\documentclass[twocolumn,showpacs,amsmath,amssymb,superscriptaddress]{revtex4-2} 
\usepackage{dcolumn,braket,color,amsmath,footnote,hyperref,bm}
\usepackage{graphicx,bm,url,longtable}
\usepackage{tabularx,multirow,braket}
\usepackage{fancybox}
\usepackage{tikz}
\usetikzlibrary{matrix}

\newcommand\+{\dagger}

\begin{document}

\title{Octupole correlations in light actinides from the interacting boson model based on the 
 Gogny energy density functional}

\author{K.~Nomura}
\email{knomura@phy.hr}
\affiliation{Department of Physics, Faculty of Science, University of Zagreb, HR-10000, Croatia}

\author{R.~Rodr\'iguez-Guzm\'an}
\author{Y.~M.~Humadi}
\affiliation{Physics Department, Kuwait University, 13060 Kuwait, Kuwait}

\author{L.~M.~Robledo}
\affiliation{Departamento de F\'\i sica Te\'orica and CIAFF, Universidad
Aut\'onoma de Madrid, E-28049 Madrid, Spain}

\affiliation{Center for Computational Simulation,
Universidad Polit\'ecnica de Madrid,
Campus de Montegancedo, Bohadilla del Monte, E-28660-Madrid, Spain
}

\author{J.~E.~Garc\'ia-Ramos}
\affiliation{Departamento de Ciencias Integradas y Centro de Estudios 
Avanzados en F\'isica, Matem\'atica y Computaci\'on, Universidad de Huelva, 
E-21071 Huelva, Spain}

\affiliation{Instituto Carlos I de F\'{\i}sica Te\'orica y Computacional,  Universidad de Granada, Fuentenueva s/n, 18071 Granada, Spain}

\date{\today}

\begin{abstract}
The quadrupole-octupole coupling and the related spectroscopic properties have been 
studied for the even-even light actinides $^{218-238}$Ra and $^{220-240}$Th. The 
Hartree-Fock-Bogoliubov approximation, based on the Gogny-D1M energy density functional, has
been employed as a microscopic input, i.e., to obtain (axially symmetric) 
mean-field potential energy surfaces as functions of the quadrupole and octupole 
deformation parameters. The mean-field potential energy surfaces have been mapped
onto the corresponding bosonic potential energy surfaces using the 
expectation value of the 
$sdf$ Interacting Boson Model (IBM) Hamiltonian  in the boson condensate state. The strength
parameters of the $sdf$-IBM Hamiltonian have been determined via this
mapping procedure. The diagonalization of the mapped IBM Hamiltonian provides 
energies for positive- and negative-parity states as well as wave functions 
which are employed to obtain transitional strengths. The results 
of the calculations compare well with available data from Coulomb excitation 
experiments and point towards a pronounced octupole collectivity 
around $^{224}$Ra and $^{226}$Th.
\end{abstract}

\maketitle

\section{Introduction}

It is a well known fact that just a handful of nuclei exhibit 
reflection asymmetric ground states with non zero octupole deformation. 
Reflection asymmetric shapes are favored in some very specific regions 
of the nuclear chart with neutron $N$ and/or proton $Z$ numbers around 
34, 56, 88, 134, $\ldots$ \cite{butler1996,butler2016}. However, 
dynamical octupole correlations have attracted considerable attention 
in recent years as they play a relevant role in the description of many 
negative parity collective states like the low-lying 1$^{-}$ states in 
the spectra of even-even nuclei that are usually considered 
fingerprints of octupole correlations \cite{robledo2011,robledo2015}. In the common situation where 
the ground state of those nuclei is quadrupole deformed, there exists a 
3$^{-}$ state, member of the corresponding negative-parity rotational 
band, which  decay through fast $E3$ transitions to the $0^{+}$ ground 
state. On the other hand, the decay of the 1$^{-}$ to the ground state 
proceeds via $E1$ transitions. The study of these as well as other  
features associated with octupole correlations, like the existence of 
alternating-parity rotational bands, has become an active research field 
with several experiments planned or already operational at  
state-of-the-art radioactive-ion beam facilities around the world. 
Within this context,  evidence of octupolarity has been found in the 
case of the light actinides ($^{220}$Rn, $^{224}$Ra and $^{222,228}$Ra 
\cite{gaffney2013,butler2020a}) and lanthanides ($^{144,146}$Ba 
\cite{bucher2016,bucher2017}). The study of octupole correlations also 
has a potential impact on other research fields. Indeed, the presence 
of static (and dynamic) nuclear octupole correlations enhance the 
fingerprints of the existence of a non-zero electric dipole moment of 
elementary particles. The existence of such an effect  would imply the 
violation of the CP symmetry implying the existence of new physics 
beyond the Standard Model of particle physics \cite{engel2013}.

From a theoretical point of view, both relativistic 
\cite{vretenar2005,niksic2011} and non-relativistic \cite{bender2003,robledo2019}
approaches rooted in the nuclear
energy density functional (EDF) framework \cite{bender2003} 
have been extensively employed to describe intrinsic nuclear
shapes and the related spectroscopic properties. In particular, the
static and dynamic aspects associated with the spontaneous breaking 
of reflection symmetry have been studied using the 
self-consistent mean-field (SCMF) approximation based on a given 
non-relativistic or relativistic EDF 
\cite{MARCOS1983,BONCHE1986,BONCHE1991,heenen1994,ROBLEDO1987,ROBLEDO1988,EGIDO1990,EGIDO1991,EGIDO1992,GARROTE1998,GARROTE1999,long2004,robledo2010,robledo2011,erler2012,robledo2012,rayner2012,robledo2013,robledo2015,bernard2016,agbemava2016,agbemava2017,xu2017,xia2017,ebata2017,rayner2020,cao2020,rayner2020gcm}. 
Dynamical beyond-mean-field correlations, stemming 
from symmetry restoration and/or fluctuations in the relevant collective 
deformations, have been considered within configuration mixing approaches
in the spirit of the Generator Coordinate Method (GCM) \cite{bender2003,robledo2019,RS}.

On the one hand,  beyond-mean-field configuration-mixing approaches 
are required to 
access  spectroscopic properties such as, the excitation energies
of negative-parity states as well as 
$B(E1)$ and $B(E3)$
reduced transition probabilities. On the other hand, beyond-mean-field approaches
become computationally expensive in medium and heavy nuclei, specially 
when several collective coordinates are to be included in the 
GCM ansatz. This drawback of the GCM justifies the introduction of computationally 
less expensive approaches like the interacting boson model (IBM)  
mapping procedures introduced in Refs. \cite{nomura2008,nomura2010}. 
In this approach, the SCMF potential 
energy surfaces (SCMF-PESs) are mapped  onto the corresponding (bosonic)
IBM-PESs as to determine some of the 
strength parameters of the corresponding IBM Hamiltonian, which is 
subsequently used to compute excitation spectra and transition probabilities. 
The method has been employed to study octupole related effects 
like the surveys of octupole related properties in the rare-earth and actinide 
regions \cite{nomura2013oct,nomura2014,nomura2015} or the 
description of octupole bands in neutron-rich odd-mass nuclei \cite{nomura2018oct}.
The SCMF-PESs have been computed using the relativistic  DD-PC1  
\cite{DDPC1} or the non-relativistic Gogny-D1M \cite{Gogny,D1M} EDFs.

Due to the renewed experimental interest in the light actinide region, 
we consider in this work the evolution of the octupole shapes 
and the resulting spectroscopic properties in a wide range of actinide
nuclei including $^{218-238}$Ra and $^{220-240}$Th. To this end, the 
quadrupole-octupole SCMF-PESs, obtained 
within the (axially symmetric) Hartree-Fock-Bogoliubov (HFB)
approximation based on the parametrization
D1M \cite{D1M} of the Gogny-EDF \cite{Gogny}, are mapped 
onto the expectation value of the interacting-boson Hamiltonian 
in the 
condensate state consisting of the monopole $L=0^+$ ($s$), quadrupole 
$2^+$ ($d$), and octupole $3^-$ ($f$) bosons \cite{IBM,engel87}.
The mapping procedure, employed to obtain the IBM-PESs from the 
SCMF-PESs, completely determines the considered quadrupole-octupole $sdf$-IBM 
Hamiltonian and its diagonalization provides  wave functions which are 
subsequently used to compute  positive- and negative-parity
spectra as well as transition strengths. 
Furthermore, by comparing with our previous spectroscopic calculations
based on the relativistic EDF DD-PC1 \cite{nomura2013oct,nomura2014}, we
demonstrate the robustness of the SCMF-to-IBM 
mapping procedure. At a qualitative (and often quantitative) level the main results and conclusions
obtained in the paper remain
the same regardless of whether relativistic or non-relativistic energy
density functional is taken as the microscopic input. Likely, this is
a consequence of both DD-PC1 and D1M being fitted to binding energies of
finite nuclei. 
In addition, the present analysis not only covers those nuclei considered
in Refs.~\cite{nomura2013oct,nomura2014}, but also explores even heavier
isotopes toward the neutron number $N=150$, in which experimental
information is not yet available. In the present study we also discuss some
quantities not covered in  Refs.~\cite{nomura2013oct,nomura2014}.

The paper is organized as
follows. The Gogny-D1M quadrupole-octupole SCMF-PESs, i.e., the microscopic
building blocks of the calculations, are discussed 
in Sec.~\ref{sec:scmf}. The mapping procedure to obtain the IBM
Hamiltonian is illustrated in Sec.~\ref{sec:ibm}. The results 
obtained for low-energy excitation spectra, electric quadrupole, octupole and dipole 
transition strengths as well as for  the transition quadrupole and octupole moments are 
discussed  
in Sec.~\ref{sec:results}. Finally, Sec.~\ref{sec:summary} is devoted to the 
concluding remarks and work perspectives.

\begin{figure*}[htb!]
\begin{center}
\includegraphics[width=\linewidth]{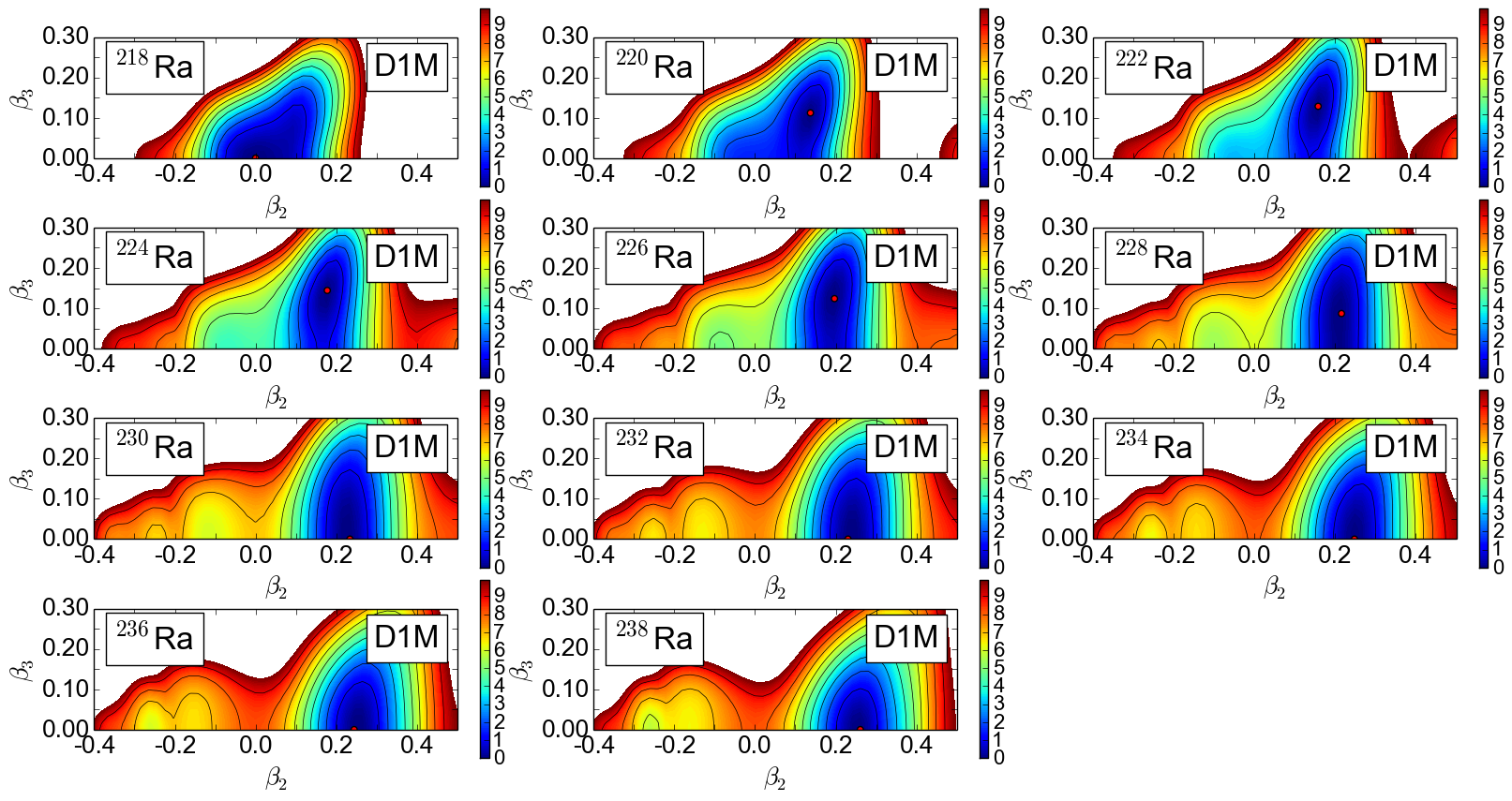}\\
\includegraphics[width=\linewidth]{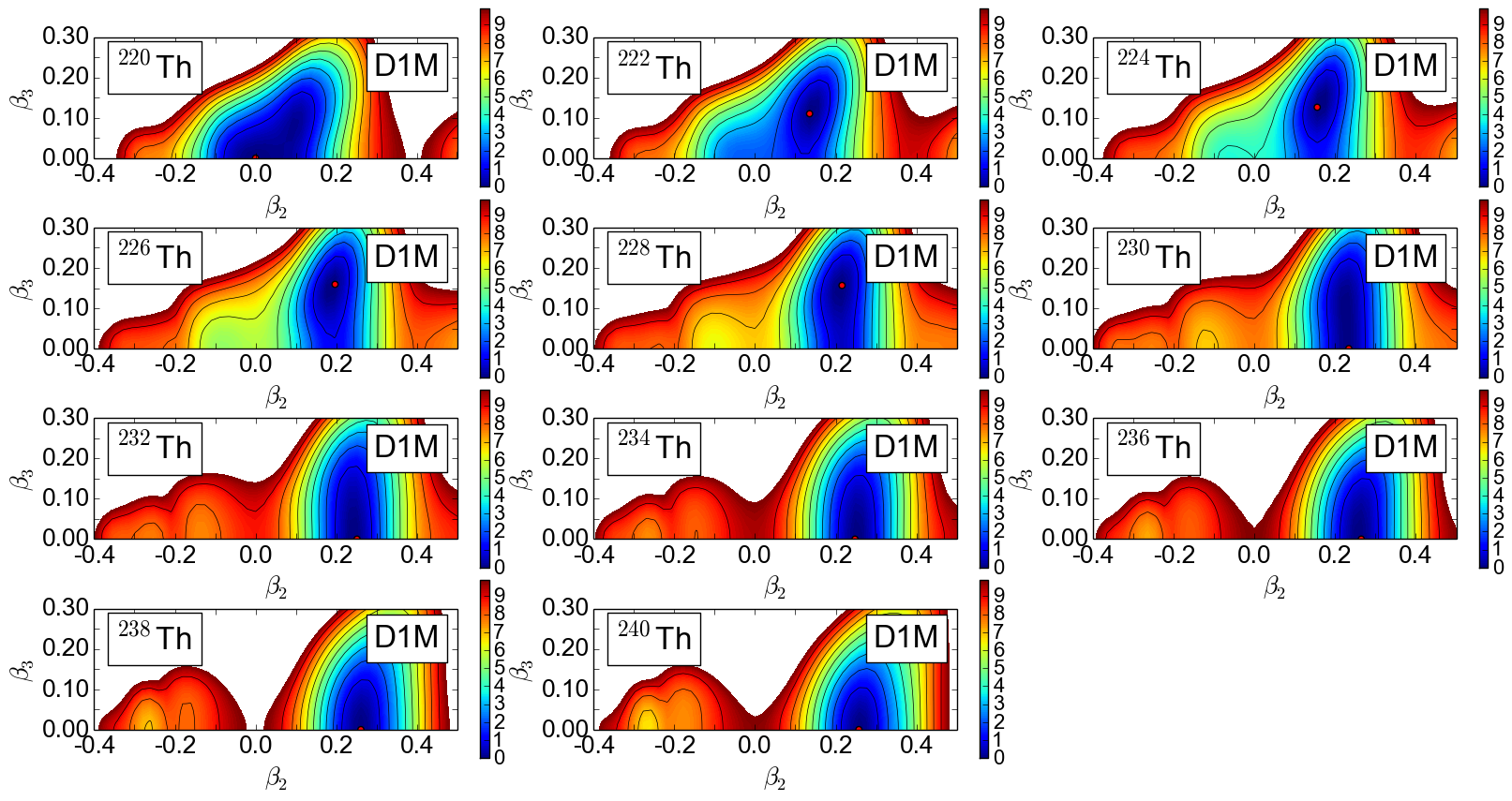}
\caption{(Color online) SCMF-PESs computed with the Gogny-D1M EDF for 
the nuclei $^{218-238}$Ra and $^{220-240}$Th.
The color code indicates the total HFB energies in MeV units, 
plotted up to 10 MeV with respect to the global minimum. For each
 nucleus, the global minimum is
 indicated by a solid circle. For more details, see the main text. } 
\label{fig:pes-gogny}
\end{center}
\end{figure*}

\section{SCMF Gogny-D1M calculations \label{sec:scmf}}

To obtain the quadrupole-octupole SCMF-PESs, the HFB equation has been solved with 
constrains on the axially symmetric quadrupole $\hat{Q}_{20}$ and octupole 
$\hat{Q}_{30}$ operators \cite{rayner2012,nomura2015}: 
\begin{align}
&\hat{Q}_{20} = z^{2} - \frac{1}{2}\left(x^{2} + y^{2}  \right) 
\nonumber\\
&\hat{Q}_{30} = z^{3} - \frac{3}{2} z\left(x^{2} + y^{2}  \right)
\end{align}
The mean values $\langle \Phi_{HFB} |\hat{Q}_{20}| \Phi_{HFB} \rangle = Q_{20}$
and $\langle \Phi_{HFB} |\hat{Q}_{30}| \Phi_{HFB} \rangle = Q_{30}$
also define the quadrupole and octupole deformation parameters
$\beta_{20}$ and $\beta_{30}$: 
\begin{align}
& \beta_{\lambda 0} = \frac{\sqrt{4 \pi (2\lambda +1)}}{3 R_{0}^{\lambda} A}  Q_{\lambda 0}
\end{align}
where $R_0=1.2 A^{1/3}$ fm. 
In the following, the subscript zero in $\beta_{\lambda 0}$'s and $Q_{\lambda 0}$'s ($\lambda=2,3$)
is omitted, unless otherwise specified. 
The center of mass is fixed at the origin to avoid spurious 
effects associated with its motion 
\cite{rayner2012,robledo2013}. The HFB quasiparticle
operators \cite{RS} have been expanded in a deformed (axially symmetric) harmonic 
oscillator (HO) basis containing 17 major shells to grant convergence for the studied
physical quantities.

The constrained  calculations 
provide a set of HFB states $| \Phi_{HFB} (\beta_{2},\beta_{3})\rangle$ 
labeled by their  
static deformation parameters $\beta_{2}$ and $\beta_{3}$ .
The HFB energies $E_{HFB}(\beta_{2},\beta_{3})$ associated
with those HFB states define the contour plots referred to as  
SCMF-PESs in this work. As the HFB energies satisfy the 
property $E_{HFB}(\beta_{2},\beta_{3}) = E_{HFB}(\beta_{2},-\beta_{3})$
only positive $\beta_{3}$ values
are considered when plotting  the SCMF-PESs.

The SCMF-PESs obtained for $^{218-238}$Ra and $^{220-240}$Th are depicted 
in Fig.~\ref{fig:pes-gogny}. Along the $\beta_{2}$-direction there
is a shape/phase transition from spherical or weakly deformed ground states
in the lightest   isotopes ($^{218}$Ra and $^{220}$Th) to well quadrupole deformed
ground states in heavier nuclei. On the other hand, the SCMF-PESs are 
rather soft along the $\beta_{3}$-direction. A global
octupole deformed 
 minimum with 
$\beta_3\approx 0.1$  already emerges for $N\approx 132$ 
($^{220}$Ra and $^{222}$Th). This minimum becomes deeper 
as one approaches the neutron number 
$N=136$ ($^{224}$Ra and $^{226}$Th). In our calculations 
the most pronounced octupole deformation effects are 
found around this neutron number with 
$\beta_3\approx 0.15$ for $^{224}$Ra and $^{226}$Th, in good agreement
with the experiment \cite{butler1996}.
Beyond this neutron number, as one moves 
towards $N=150$, the corresponding  
$\beta_3$ values decrease and reflection
symmetric HFB ground states are obtained
for the heaviest isotopes in both chains. 

Previous SCMF calculations including the 
quadrupole and octupole constrains simultaneously
can be found in the literature for nuclei
in this region of the nuclear chart. For example, 
calculations have been carried out in 
Ref.~\cite{nomura2014} for $^{218-228}$Ra and $^{220-232}$Th
using the relativistic 
DD-PC1 EDF \cite{DDPC1}. The overall
systematic of the quadrupole  and octupole deformations 
associated with the DD-PC1 SCMF-PESs is similar to the 
one obtained in the present study with the Gogny-D1M EDF.
However, in the case 
of the DD-PC1 EDF, the $N=132$ isotopes ($^{220}$Ra and $^{222}$Th)
exhibit a reflection symmetric SCMF ground state
while those nuclei are predicted to be octupole deformed in the 
Gogny-D1M calculations. Pronounced octupole deformation effects
are predicted by both EDFs for $^{224}$Ra and $^{226}$Th though
deeper global minima are found in the relativistic approach. 
The quadrupole-octupole coupling has  been studied for Rn, Ra and 
Th nuclei in Ref.~\cite{robledo2013}. A comparison 
of several relativistic EDFs in a survey of octupole correlations
can be found in Ref.~\cite{agbemava2016}. A thorough account over a 
large set of even-even nuclei of observables associated to octupole correlations   
was presented in Refs.~\cite{robledo2011,robledo2015} using the 
Gogny-HFB approach, parity projection and octupole 
configuration mixing. Octupole deformations have also
been studied for Ra isotopes \cite{robledo2010} using the HFB
approach  based 
on the Barcelona-Catania-Paris (BCP) \cite{robledo2010} 
and Gogny-D1S \cite{D1S} EDFs.

%
%
\begin{figure*}[htb!]
\begin{center}
\includegraphics[width=\linewidth]{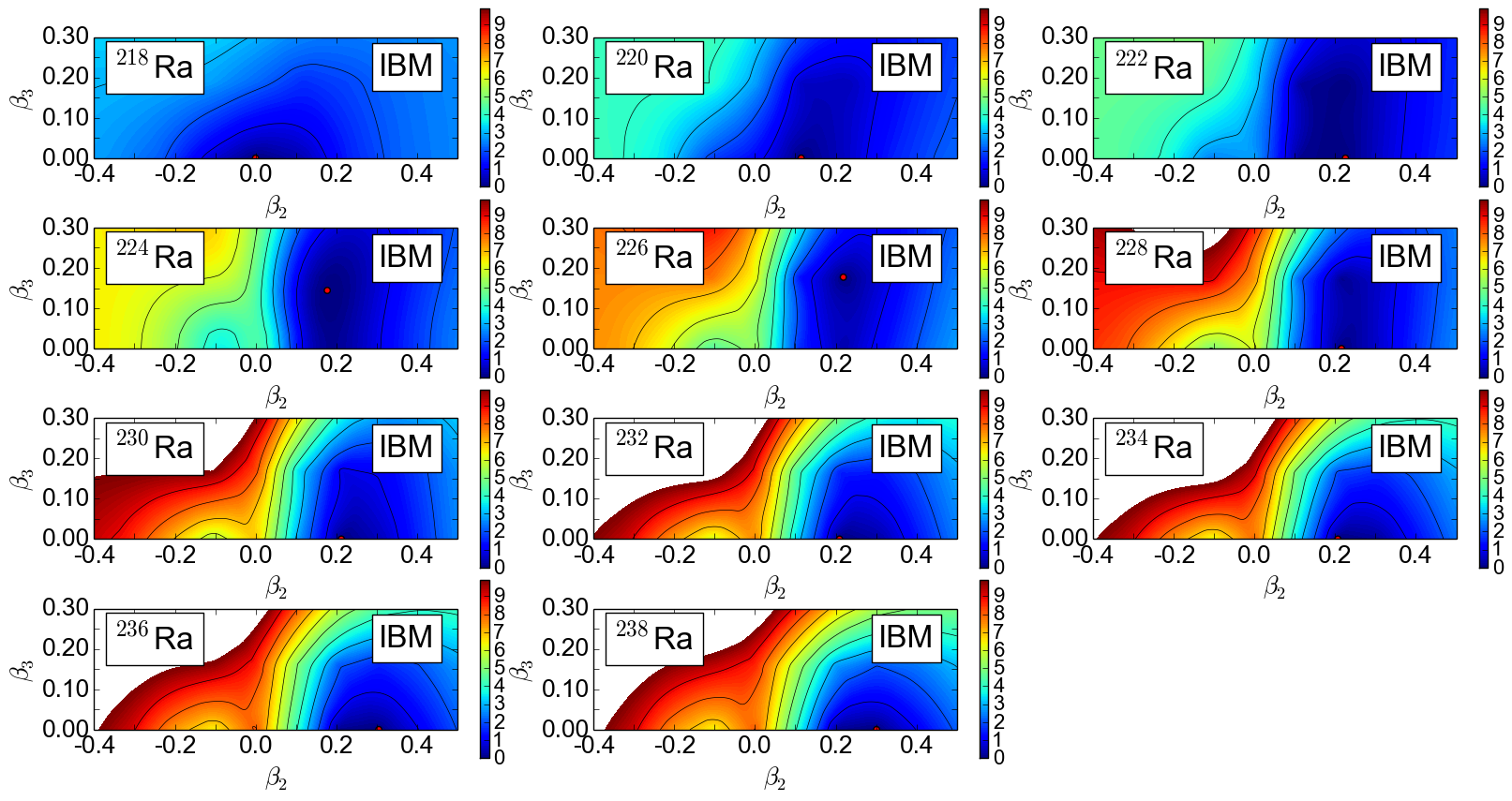}\\
\includegraphics[width=\linewidth]{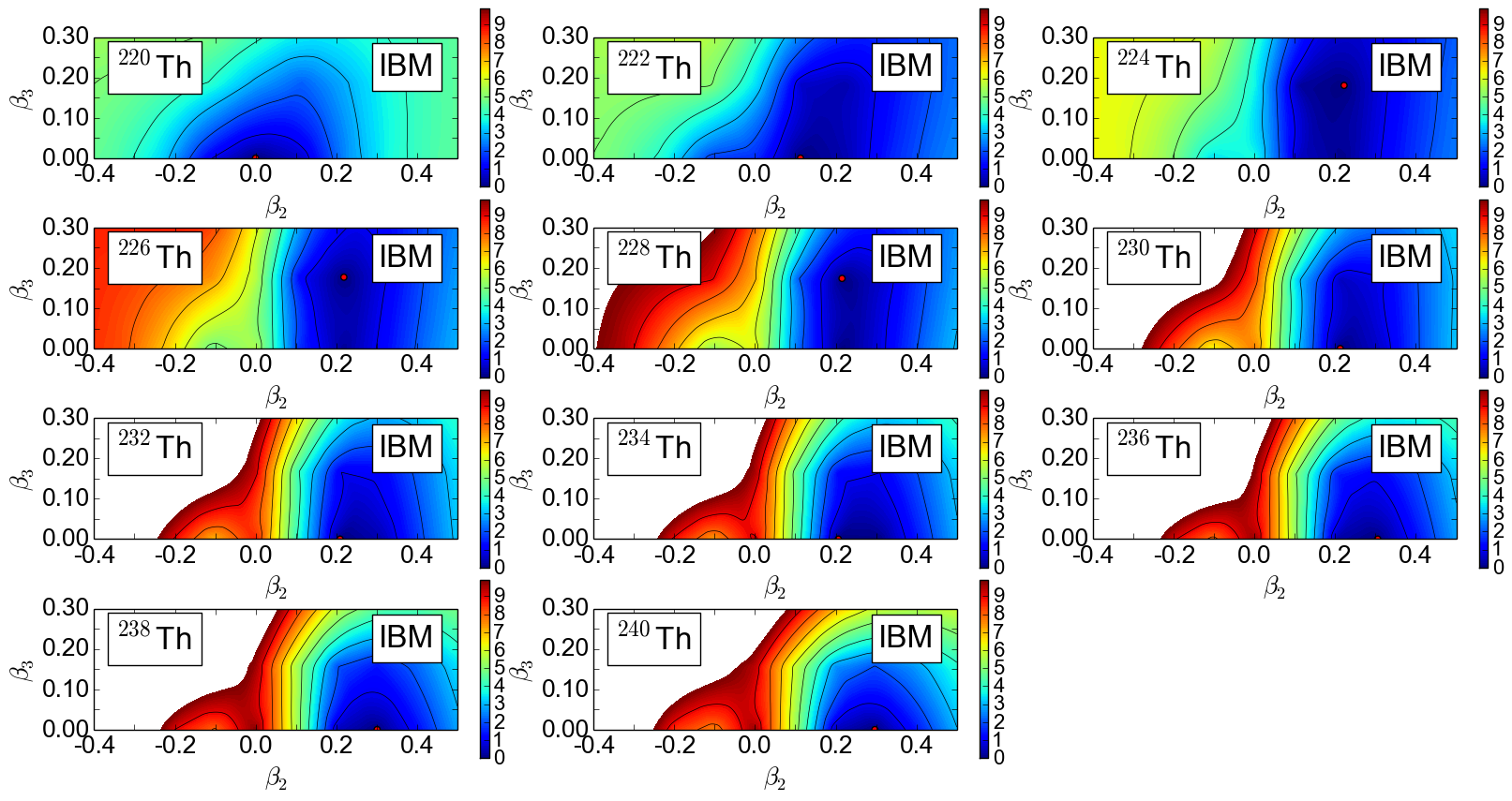}
\caption{(Color online) IBM-PESs computed with the mapped $sdf$-IBM Hamiltonian  Eq.~(\ref{eq:ham})
for 
the nuclei $^{218-238}$Ra and $^{220-240}$Th. For more details, see the main text.
} 
\label{fig:pes-ibm}
\end{center}
\end{figure*}

%
%
\begin{figure}[htb!]
\begin{center}
\includegraphics[width=\linewidth]{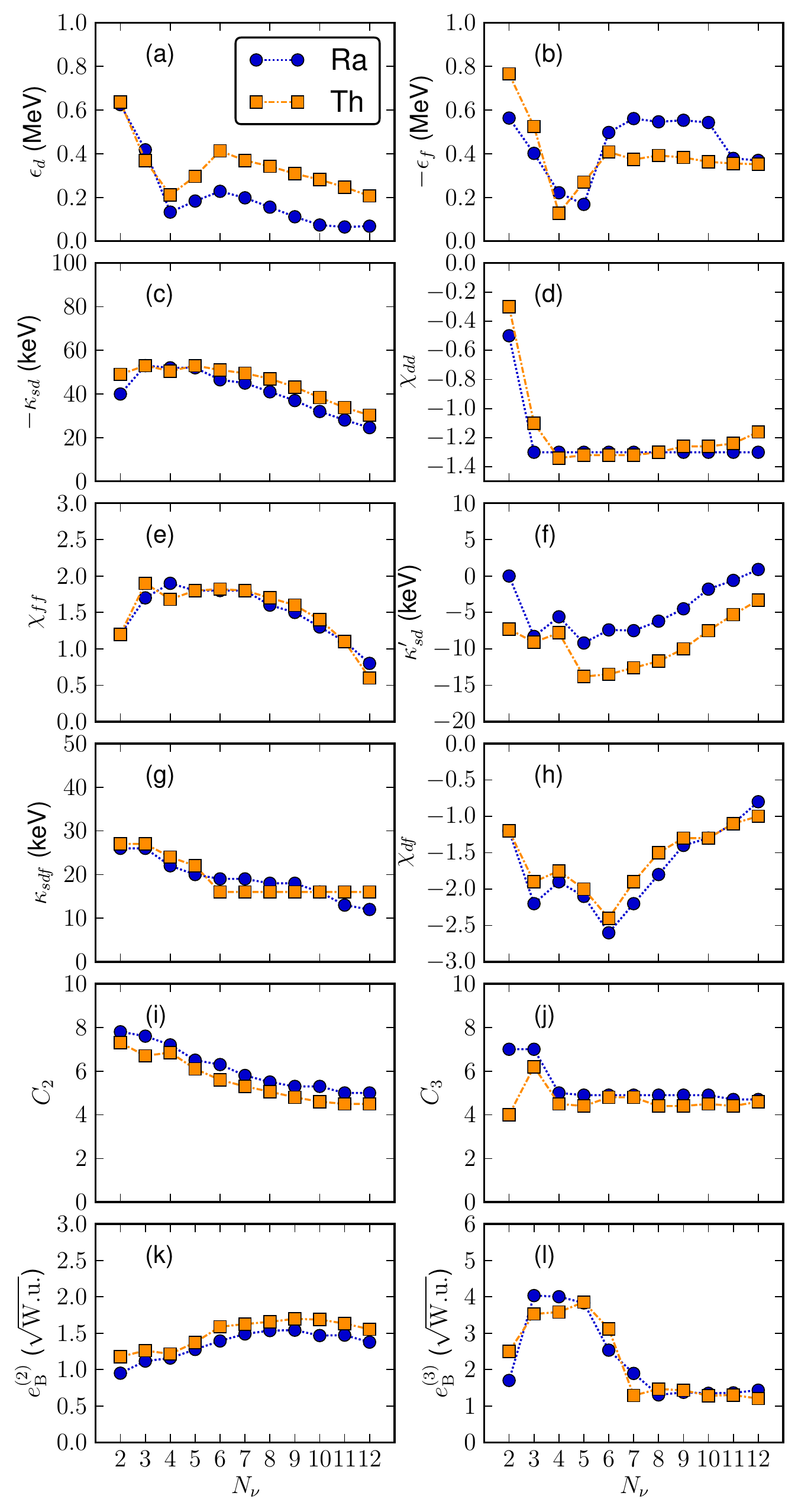}
\caption{(Color online) The strength parameters of the $sdf$-IBM 
Hamiltonian Eqs.~(\ref{eq:sd}) to (\ref{eq:q3}) (panels (a) to (h)), 
the coefficients $C_\lambda$'s for the deformation parameters (panels (i) and (j)), 
and the boson effective charges $e_B^{(\lambda)}$ 
Eqs.~(\ref{eq:eb-1})
 and (\ref{eq:eb-2}) (panels (k) and (l))
are plotted as functions of the neutron boson number $N_\nu$ for 
$^{218-238}$Ra and $^{220-240}$Th. } 
\label{fig:parameter}
\end{center}
\end{figure}

\section{Mapping onto the boson system\label{sec:ibm}}

Having the (fermionic)  Gogny-D1M SCMF-PESs at hand, we map them onto 
the corresponding (bosonic) IBM-PESs using the methods developed 
in Refs.~\cite{nomura2008,nomura2010,nomura2013oct}. In order to account 
for negative-parity states the IBM space includes, in addition to the positive-parity 
monopole $s$ ($L=0^+$) and quadrupole $d$ ($L=2^+$) bosons, the 
negative-parity 
$f$ ($L=3^-$) boson. Within the IBM framework, bosons represent collective 
pairs of valence nucleons \cite{OAI}. In particular, the $f$ boson can be viewed
as formed  by coupling the normal and 
unique parity orbitals  $\pi (i_{13/2}\otimes f_{7/2})^{(3^-)}$ 
and $\nu (j_{15/2}\otimes g_{9/2})^{(3^-)}$ in the light actinides 
with $Z\approx 88$ and $N\approx 136$. In the  
usual $sdf$-IBM phenomenology, the number of $f$ bosons 
involved in the IBM space is  limited to one or, at most, three. 
In the present work, we do not assume any such truncation for the 
$f$-boson number. Thus, the numbers $n_s$, $n_d$ and $n_f$ of $s$, $d$, and $f$ bosons
are arbitrary and satisfy the condition  that the total 
boson number 
$N_\mathrm{B}=n_s + n_d + n_f$ is conserved for  a given nucleus.

The mapping of the Gogny-D1M $(\beta_{2},\beta_{3})$-PESs onto
the IBM ones is achieved by introducing the 
intrinsic  state for the boson system 
\cite{ginocchio1980}: 
\begin{align}
 \label{eq:coherent}
|\phi\rangle=\frac{1}{\sqrt{N_\mathrm{B}!}}(b_c^{\dagger})^{N_\mathrm{B}}\ket{0}, 
\end{align}
where $N_{B}$ and $\ket{0}$ denote the number of bosons and the boson 
vacuum, respectively. The condensate boson operator $b_c^\+$ is given by 
\begin{align}
\label{eq:bc}
b_c^\+=(1+\alpha_2^2+\alpha_3^2)^{-1/2}
(s^\+ +\alpha_{2}d_0^\+ + \alpha_{3}f_{0}^\+), 
\end{align}
with amplitudes $\alpha_{2}$ and $\alpha_{3}$. The doubly-magic nucleus
$^{208}$Pb is taken as boson vacuum. Therefore, $N_\mathrm{B}$ 
runs from 5 to 15  (6 to 16) for  $^{218-238}$Ra ($^{220-240}$Th).
The amplitudes $\alpha_{2}$ and $\alpha_{3}$ can be related 
to the 
deformation parameters $\beta_2$ and $\beta_3$
as $\alpha_{2}=C_2\beta_2$ and $\alpha_{3}=C_3\beta_3$
\cite{ginocchio1980,nomura2014,nomura2015} where, $C_2$ and $C_3$  represent 
dimensionless  parameters.

The IBM-PES is  obtained analytically, by taking the expectation 
value of the $sdf$-IBM Hamiltonian in the 
boson condensate state  Eq.~(\ref{eq:coherent}). The 
$sdf$-IBM Hamiltonian  is the sum of the 
Hamiltonians for the $sd$ and $f$ boson 
spaces plus a coupling  $\hat H_{sdf}$ between them:
\begin{align}
 \label{eq:ham}
\hat H= \hat H_{sd} + \hat H_{f} + \hat H_{sdf}. 
\end{align}

The $sd$-boson Hamiltonian reads
\begin{align}
 \label{eq:sd}
\hat H_{sd} = 
\epsilon_d\hat n_{d} + \kappa_{sd}\hat Q_{sd}\cdot\hat Q_{sd}+\kappa_{sd}'\hat
L_d\cdot\hat L_d, 
\end{align}
where the first term represents the number operator for
the $d$ bosons with $\epsilon_d$ being the single
$d$ boson energy relative to the $s$ boson one. 
The second term represents the quadrupole-quadrupole interaction with 
strength $\kappa_{sd}$ and the quadrupole operator 
$\hat Q_{sd}=s^{\dagger}\tilde d+d^{\dagger}\tilde s+\chi_{dd}[d^{\dagger}\times\tilde
  d]^{(2)}$. The third term in Eq.~(\ref{eq:sd}) is the 
rotational term with the angular 
momentum operator $\hat L_d=\sqrt{10}[d^{\dagger}\times\tilde d]^{(1)}$.

The Hamiltonian for the $f$-boson space reads 
\begin{align}
 \label{eq:f}
\hat H_f =\epsilon_f\hat
n_f+\kappa_f\hat Q_{f}\cdot\hat Q_{f}+\kappa_f'
\hat L_f\cdot\hat L_f,
\end{align}
with the $f$-boson quadrupole operator $\hat Q_f$ and angular
momentum operator $\hat L_f$ being 
$\hat Q_{f}=[f^{\dagger}\times\tilde f]^{(2)}$
and $\hat L_f = \sqrt{28}[f^\+\times\tilde f]^{(1)}$, respectively.

The $sdf$ Hamiltonian employed here takes the following form: 
\begin{align}
 \label{eq:sdf}
\hat H_{sdf} = \kappa_{sdf}'\hat Q_{sd}\cdot\hat Q_{f}+\kappa_{sdf}\hat O\cdot\hat O, 
\end{align}
The last term in Eq.~(\ref{eq:sdf}) is the octupole-octupole 
interaction with the strength parameter $\kappa_{sdf}$. The octupole 
operator takes the form 
\begin{align}
\label{eq:q3}
 \hat O=s^{\dagger}\tilde f+f^{\dagger}\tilde
  s+\chi_{df}[d^{\dagger}\times\tilde
  f+f^{\dagger}\times\tilde
  d]^{(3)}, 
\end{align}
with $\chi_{df}$ being a parameter. 

For convenience, we introduce a new parameter $\chi_{ff}$ so that
$\kappa_{sdf}' = 2\kappa_{sd}\chi_{ff}$, and 
$\kappa_f = \kappa_{sd}\chi_{ff}^2$, and assume  $\kappa'_f =
\kappa_{sd}'$. 
The independent parameters of the Hamiltonian are, therefore, 
$\epsilon_d$, $\epsilon_f$, $\kappa_{sd}$, $\kappa'_{sd}$, $\chi_{dd}$, $\chi_{ff}$, 
$\kappa_{sdf}$, and $\chi_{df}$ as well as the  coefficients $C_2$ and $C_3$
for the $\beta_2$ and $\beta_3$ deformations. These parameters 
are determined via the mapping procedure. 
The Hamiltonian in Eq.~(\ref{eq:ham}) 
is similar to the one employed
in our previous study in the  rare-earth region \cite{nomura2015}, 
except for the $\hat L_f\cdot\hat L_f$ term  considered 
in  this work. This 
rotational correction term is considered 
because a good amount of $f$-boson components 
is present in the calculated yrast states for both parities
and the inclusion of this term 
has a sizable effect on the moments 
of inertia obtained for the positive and negative-parity yrast bands.
A more detailed account of the 
other terms  as well as the analytical form of the IBM-PES 
as a function of the $\beta_{2}$ and $\beta_{3}$ deformations
can be found in Ref.~\cite{nomura2015}.

The strength parameters of the 
$sdf$-IBM Hamiltonian in Eq.~(\ref{eq:ham})
are determined so that the IBM-PES  reproduces the 
topology of the Gogny-D1M SCMF-PES 
around the global minimum. 
Since the Hamiltonian (\ref{eq:ham}) contains a large number of
parameters, an unconstrained fitting can land in local minima far away from
the physical solution. Therefore, it is always convenient to fit 
the parameters in a controlled, physically inspired, way by using the following
procedure: First, the strength parameters of the
$sd$-boson space Hamiltonian  $\hat H_{sd}$ (\ref{eq:sd})  
($\epsilon_d$, $\kappa_{sd}$, $\chi_{dd}$, and $C_{2}$) are fixed by 
carrying out the mapping along the $\beta_3=0$ axis in such a way that the
curvature in $\beta_2$ around the absolute (prolate) minimum, depth of the potential well, and
the energy difference between the prolate and oblate minima are reproduced. 
Only the parameter $\kappa_{sd}'$ for the $\hat L_d\cdot\hat L_d$ term 
in Eq.~(\ref{eq:sd}) is determined independently in such a way 
that the bosonic 
cranking moment of inertia (see Ref.~\cite{nomura2011rot} for details) 
at the absolute minimum along the $\beta_3=0$ axis 
matches the Thouless-Valatin \cite{TV} 
moment of inertia for the $2^+_1$ state computed with the reflection symmetric SCMF cranking model. 
Second, the strengths parameters related to the $f$-boson space
Hamiltonian $\hat H_{f}$ \ref{eq:f}  and the ones
related to the coupling between $sd$- and $f$-boson spaces $\hat H_{sdf}$ \ref{eq:sdf}
($\epsilon_f$, $\chi_{ff}$, $\kappa_{sdf}$, $\chi_{df}$, and $C_{3}$)
are determined in such a way that the following features of the SCMF-PES in the
$(\beta_2,\beta_3)$ space are reproduced as closely as possible: 
curvatures along the $\beta_2$ and $\beta_3$ directions around the global minimum,
location of the minimum, and steepness of the potential both in $\beta_2$
and $\beta_3$ directions. 

To uniquely determine the parameters, the following constraints are also
considered, so as to be more or less consistent with  our previous
results \cite{nomura2013oct,nomura2014,nomura2015} and earlier
phenomenological studies within the $sdf$-IBM framework on the same mass
region (e.g., Refs.~\cite{zamfir2001,zamfir2003}): 
(i) each parameter should evolve gradually with boson number; (ii) since
most of the considered nuclei are strongly quadrupole deformed, the
parameter $\chi_{dd}$ should take a value close to the one in the
SU(3) limit of the IBM $\chi_{dd}=-1.32$ \cite{IBM}; (iii) $d$-boson energy 
$\epsilon_d$ should decrease with boson number; (iv) $\epsilon_d$ should
be lower in magnitude than the $f$-boson energy $\epsilon_f$, except for
the strongly octupole deformed nuclei around $N=136$; 
(v) the strengths $\kappa_{sd}$ and $\kappa_{sdf}$ should decrease in
magnitude as the boson number increases.

The  mapped $sdf$-IBM-PESs  are depicted 
in Fig.~\ref{fig:pes-ibm} for the studied nuclei. As expected, 
the original Gogny-D1M $(\beta_2,\beta_3)$-PESs are nicely reproduced 
around the global minimum. The IBM-PESs  are, however, much flatter 
far away from this minimum. This is a common feature  found in previous IBM studies 
and can be attributed  to the size of the IBM model space \cite{ginocchio1980,nomura2008}.
The boson configuration space consists of only valence 
nucleons while all the nucleons are involved in the  Gogny-HFB calculation. 
The resulting $sdf$-IBM Hamiltonian, with the 
strength parameters  determined via the mapping procedure, is then diagonalized to 
obtain excitation energies and transition strengths for 
a given nucleus.

The strength parameters obtained for the $sdf$-IBM 
Hamiltonian are plotted in 
panels (a) to (j) of 
Fig.~\ref{fig:parameter} as functions of the 
neutron boson number $N_\nu$, which equals $N_\mathrm{B}-(Z-82)/2$. 
The boson-number dependence of the parameters along an 
isotopic chain reflects the corresponding  structural changes. 
Most of the parameters are smooth functions of $N_\nu$. 
However, some of the parameters for the interaction terms 
involving  $f$ bosons, e.g., $\epsilon_{f}$, and $\chi_{df}$, display 
abrupt changes  around $N=136$. This results from the 
difference in the topology of the SCMF-PESs
corresponding to neighboring isotopes in this 
transitional region (see, Fig.~\ref{fig:pes-gogny}).

%
%
\begin{figure*}[htb!]
\begin{center}
\includegraphics[width=0.6\linewidth]{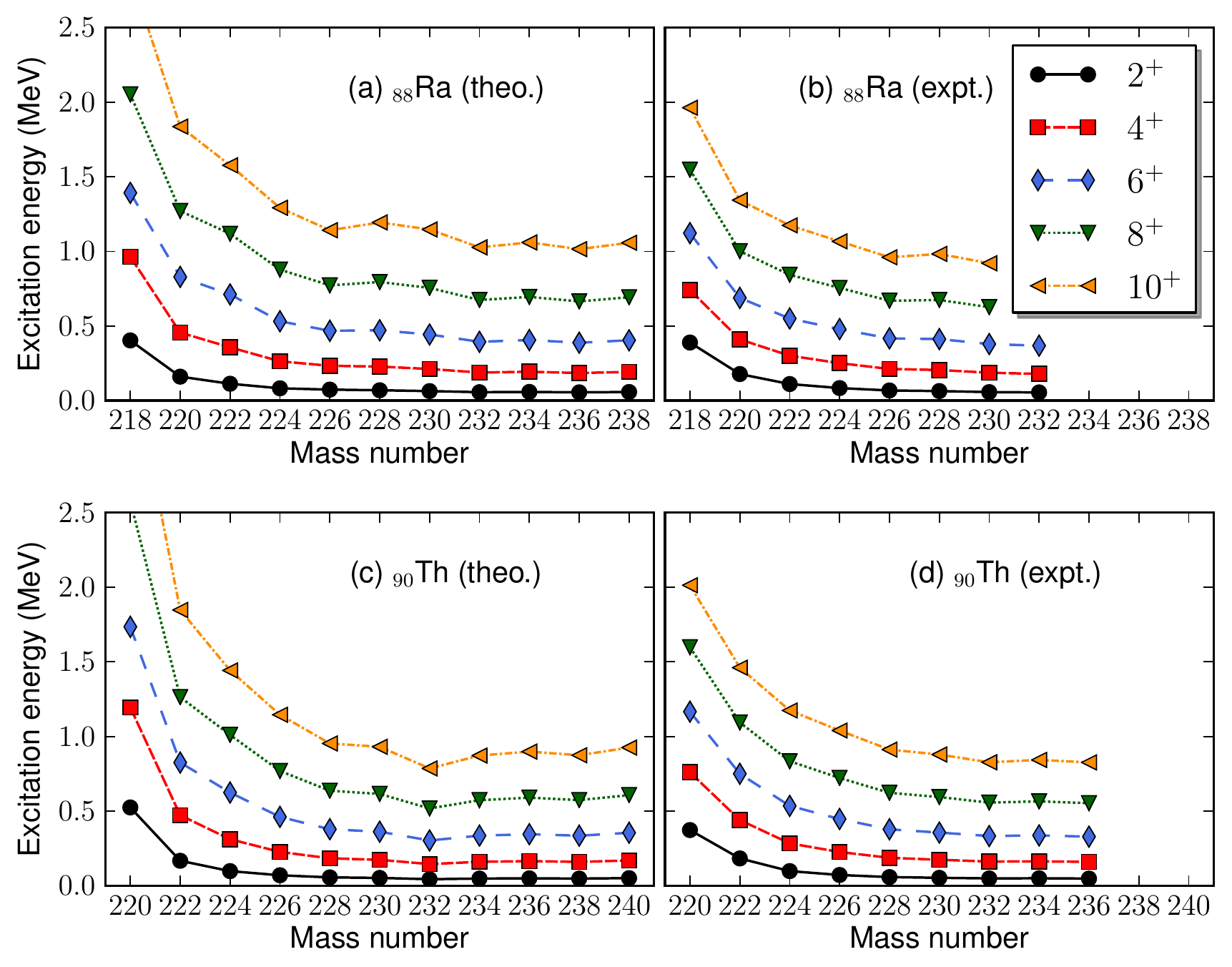}\\
\includegraphics[width=0.6\linewidth]{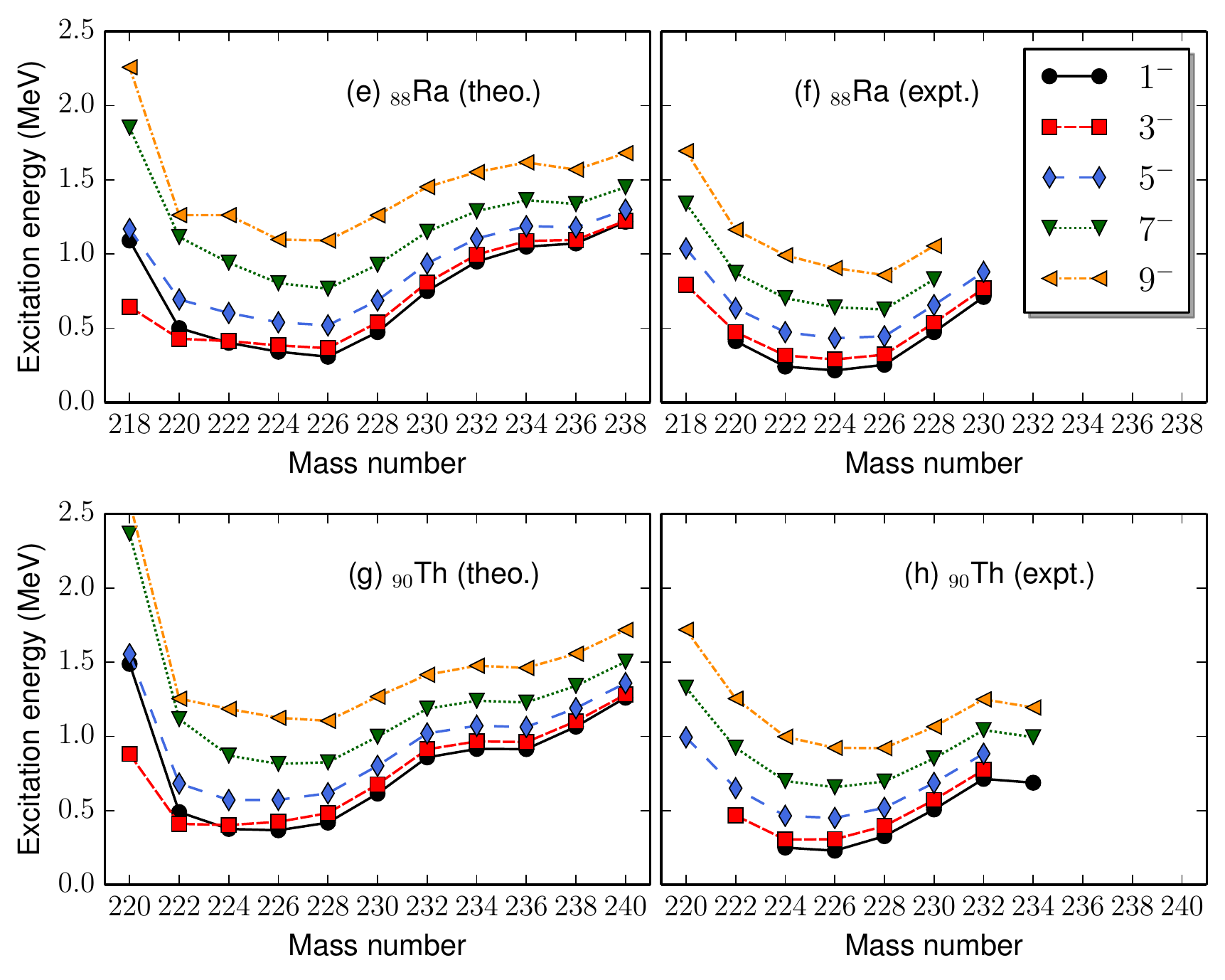}
\caption{(Color online) 
Low-energy even-spin positive and odd-spin negative-parity 
excitation spectra of yrast states 
for  $^{218-238}$Ra and $^{220-240}$Th 
computed with the $sdf$-IBM Hamiltonian  Eq.~(\ref{eq:ham}). 
Experimental data are taken from Ref.~\cite{data}.} 
\label{fig:level}
\end{center}
\end{figure*}

%
%
\begin{figure}[htb!]
\begin{center}
\includegraphics[width=\linewidth]{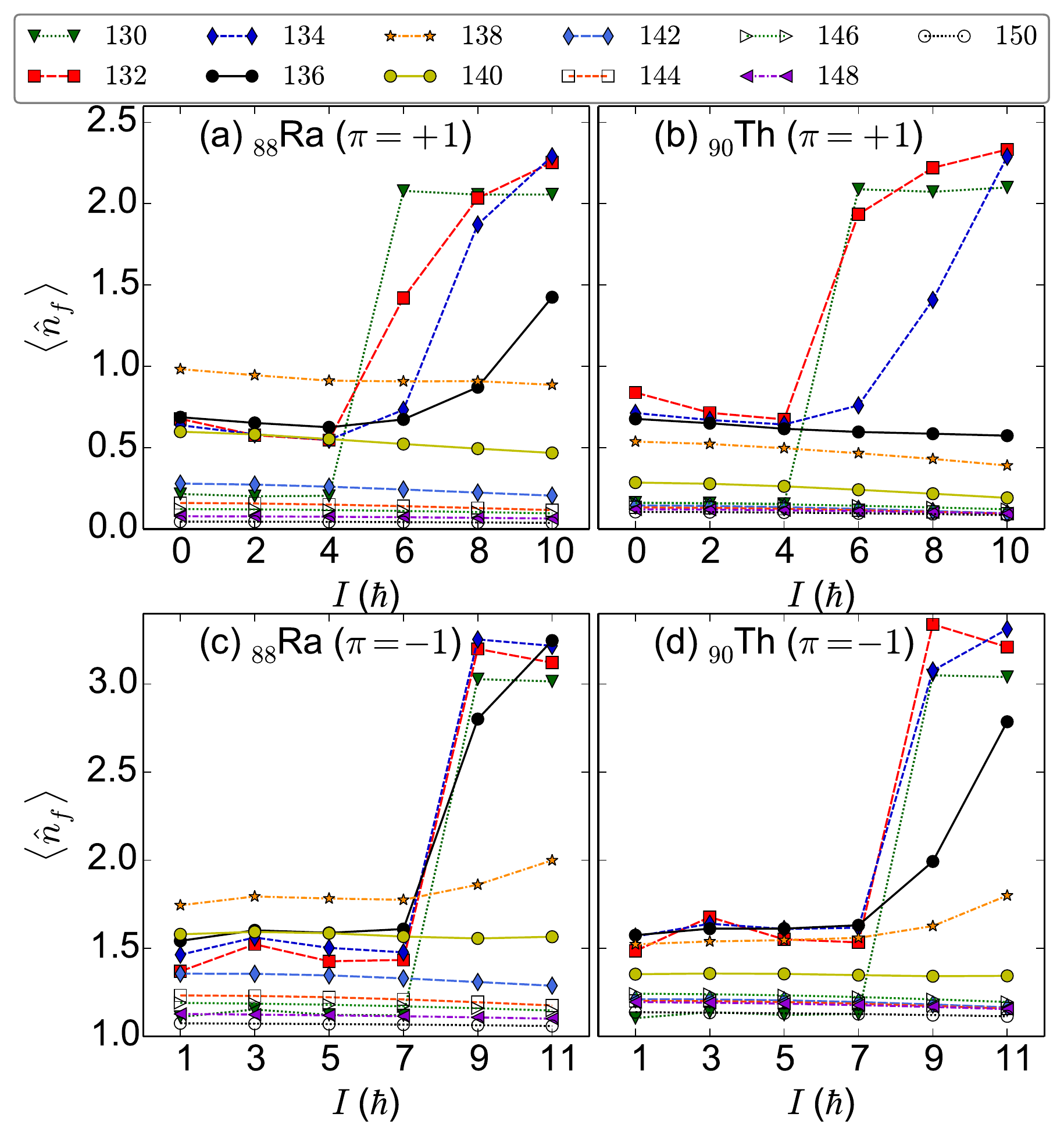}
\caption{(Color online) 
The $f$-boson contents in the wave functions of the 
even-spin positive-parity (a,b) and odd-spin negative-parity 
yrast states (c,d) in $^{218-238}$Ra and $^{220-240}$Th, obtained as 
the expectation value  $\braket{\hat n_f}$ in a given state, are plotted
as functions of the spin $I$. Different symbols denote the calculated
 quantities $\braket{\hat n_f}$ for the considered nuclei (their neutron
 numbers are indicated in the legend on the top), and are connected by lines.} 
\label{fig:wf}
\end{center}
\end{figure}

%
%
\begin{figure}[htb!]
\begin{center}
\includegraphics[width=\linewidth]{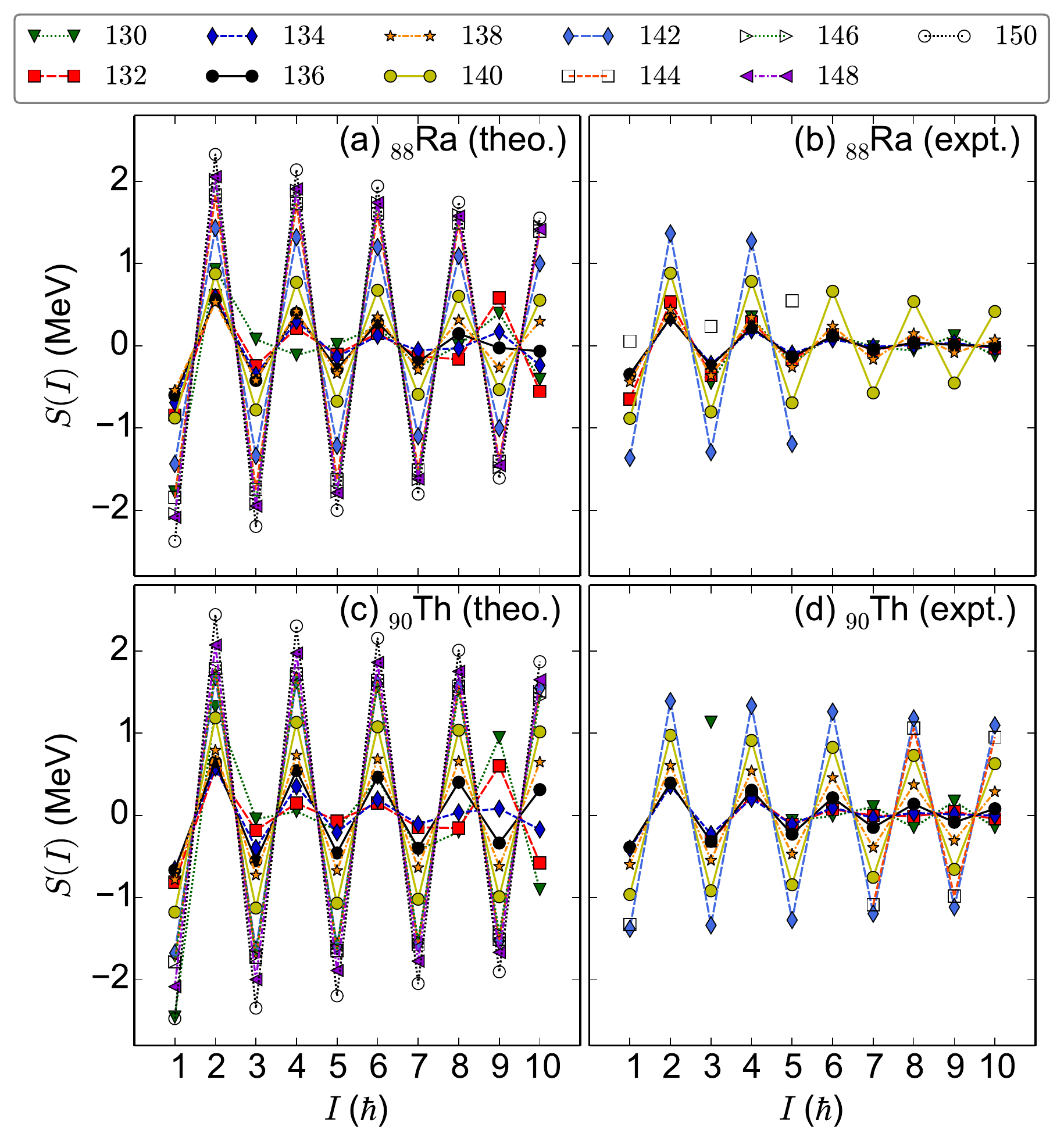}
\caption{(Color online) 
The relative energy splitting between 
positive- and negative-parity
yrast bands $S(I)$ Eq.(\ref{alter-quant}), obtained for 
$^{218-238}$Ra and $^{220-240}$Th, is plotted as a function 
of the  spin $I$. The legend on the top indicates the neutron
 numbers for those nuclei that are plotted in each panel.} 
\label{fig:altern}
\end{center}
\end{figure}

%
%
\begin{figure}[htb!]
\begin{center}
\includegraphics[width=\linewidth]{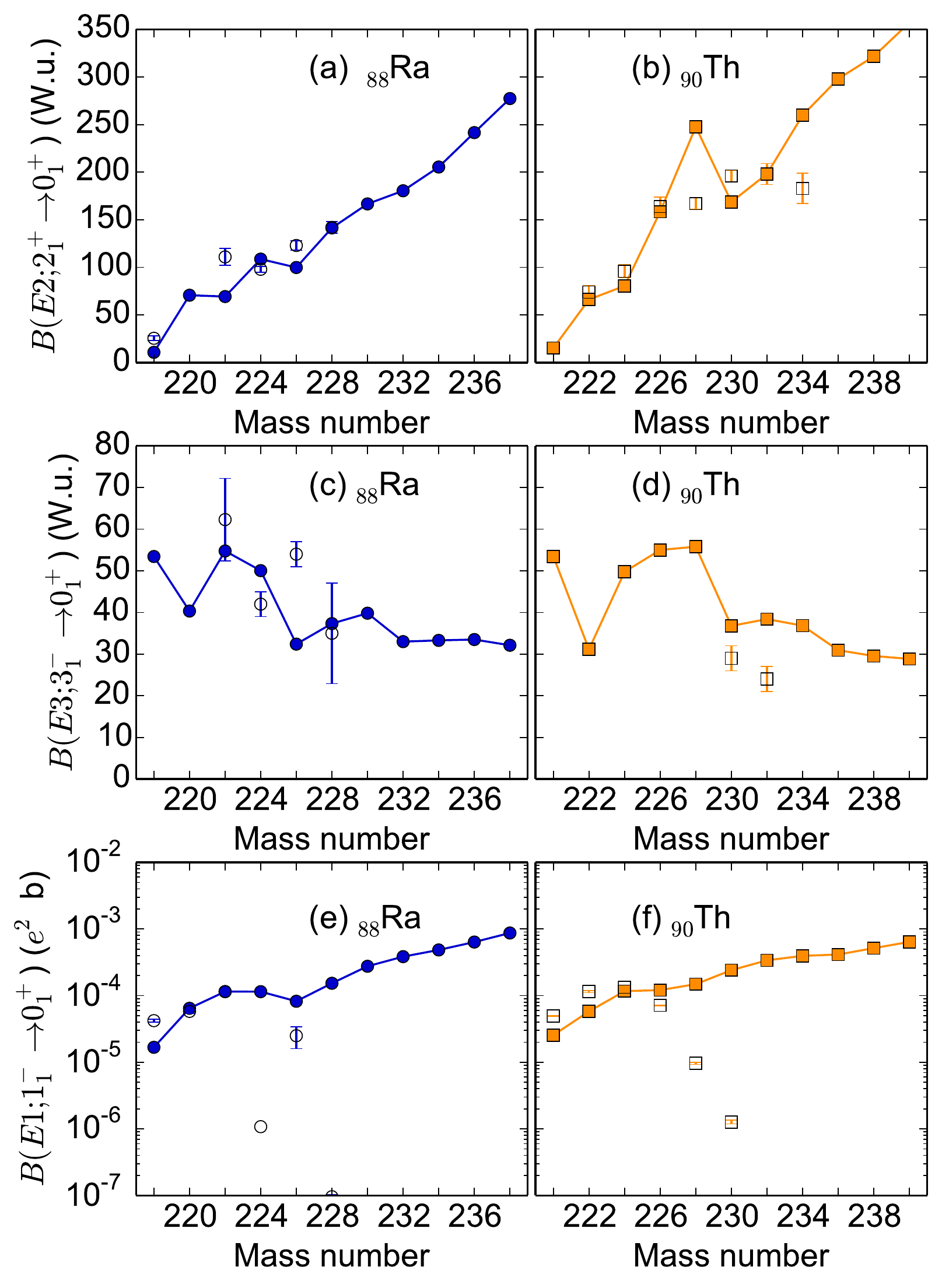}
\caption{(Color online) Reduced transition probabilities
$B$(E2; $2^+_1\to 0^+_1$) (a,b), 
$B$(E3; $3^-_1\to 0^+_1$) (c,d), 
and $B$(E1; $1^-_1\to 0^+_1$) (e,f) 
for $^{218-238}$Ra and $^{220-240}$Th. Theoretical
values are represented by filled symbols connected 
by lines. Experimental data have been taken from 
Refs.~\cite{gaffney2013,butler2020a,data}. They are  
represented by open symbols with error bars. 
The $B$(E2) and $B$(E3) rates are in Weisskopf units
while the  $B$(E1) rates in $e^2\cdot$b units
are plotted using a  logarithmic scale.
} 
\label{fig:transitions}
\end{center}
\end{figure}

%
%
\begin{figure*}[htb!]
\begin{center}
\includegraphics[width=0.7\linewidth]{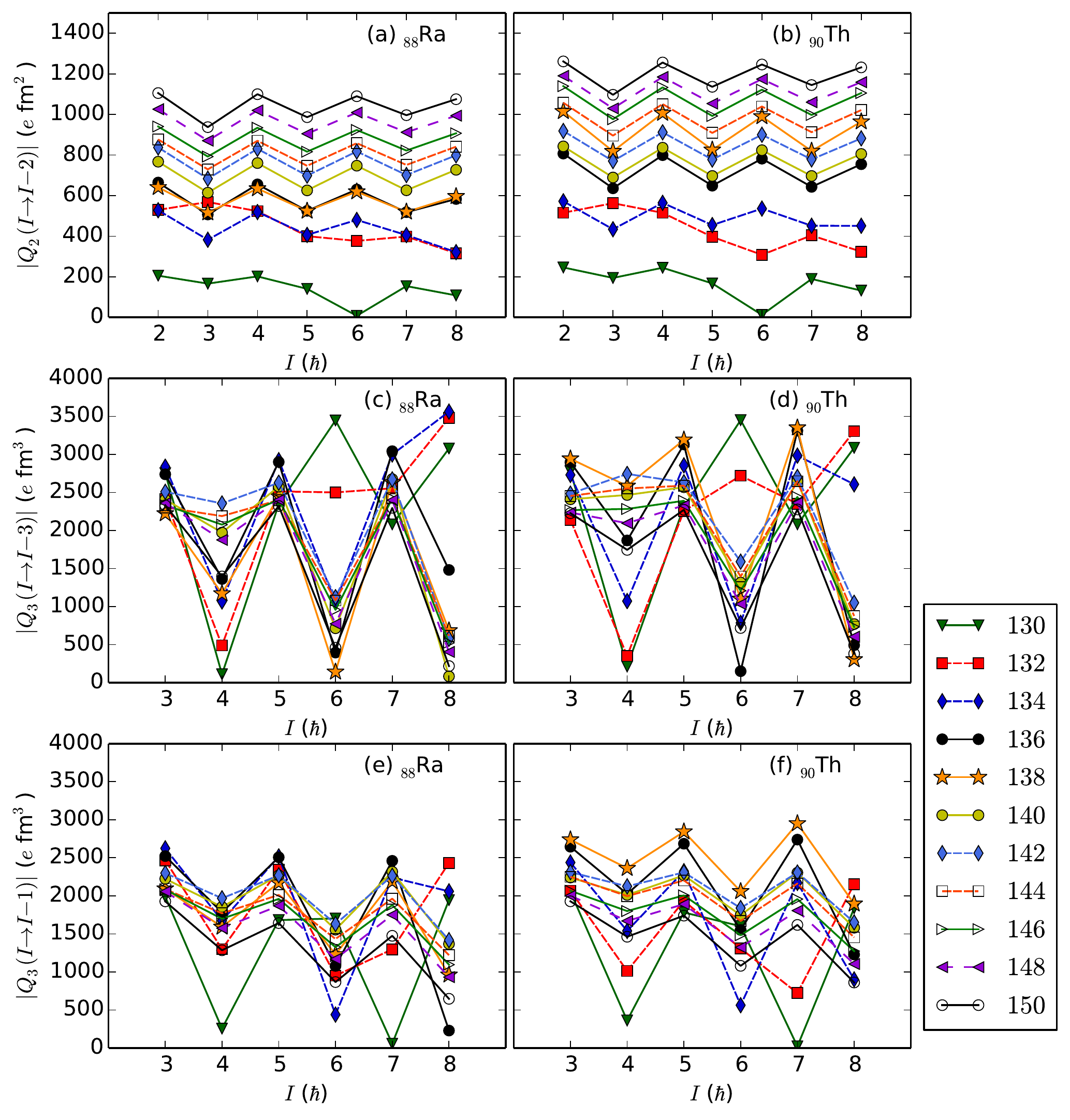}
\caption{(Color online) 
The transition quadrupole and octupole moments 
(in $e\cdot$fm$^{\lambda}$ units) 
obtained for $^{218-238}$Ra and $^{220-240}$Th
are plotted as functions of the spin $I$. Different symbols represent the calculated quadrupole and octupole moments for the
 considered nuclei (their neutron numbers are indicated in the legend on
 the right), and are connected by
 lines. For more details, see the main text.
} 
\label{fig:mom}
\end{center}
\end{figure*}

%
%
\begin{figure}[htb!]
\begin{center}
\includegraphics[width=\linewidth]{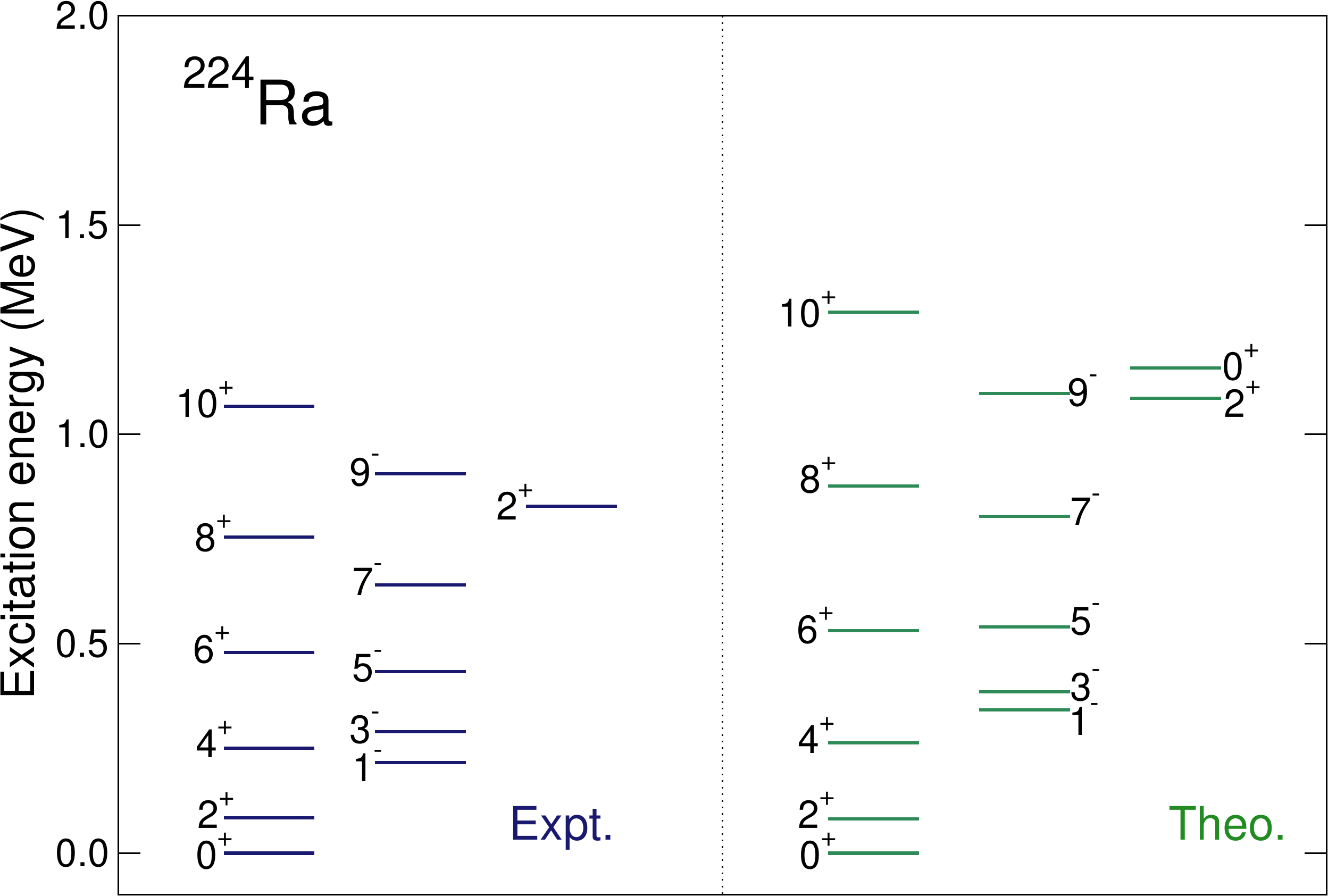}
\includegraphics[width=\linewidth]{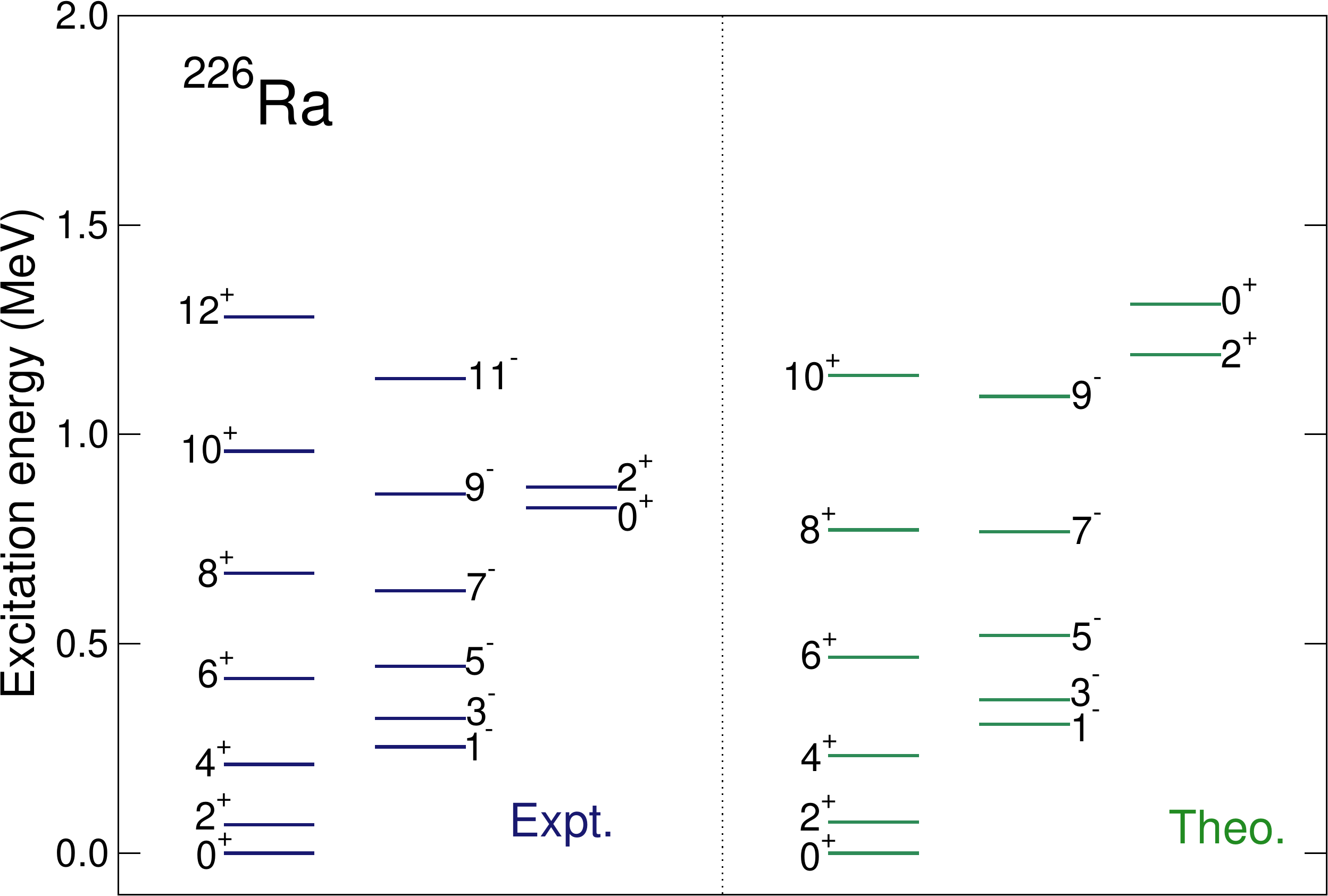}
\caption{(Color online) The energy spectra obtained for 
$^{224}$Ra (top panel) and $^{226}$Ra (bottom panel)
are compared with the experimental ones \cite{gaffney2013}.
} 
\label{fig:ra224}
\end{center}
\end{figure}

\section{Spectroscopic properties\label{sec:results}}

\subsection{Systematic of excitation spectra}

The low-energy excitation spectra corresponding 
to even-spin positive and odd-spin negative-parity yrast states
are plotted in Fig.~\ref{fig:level} as functions of the mass 
number $A$. Those states are assumed to be 
members of the $K^\pi=0^+_1$ and $0^-_1$ bands. The excitation energies
of the positive-parity states decrease with increasing neutron  
number. This reflects
the onset of pronounced quadrupole deformation effects with increasing 
neutron number (see, Figs.~\ref{fig:pes-gogny} and \ref{fig:pes-ibm})
and the corresponding transition from vibrational to 
well-developed rotational bands. For both isotopic chains, the predicted 
positive-parity spectra agree reasonably well with the
experimental ones also included in the figure.

The excitation energies of the negative-parity states exhibit a 
parabolic behavior as functions of the neutron number. The lowest
 excitation energies
correspond to $N \approx 136$ isotopes. Around this 
neutron number the predicted negative-parity band
lies quite close in energy to the positive-parity band. This situation 
corresponds to an alternating-parity rotational band 
(see, Sec.~\ref{alternating-parity-band})
that is a neat fingerprint of
permanent octupole deformation \cite{butler1996}. For larger
neutron numbers, the negative-parity band is higher in energy
and completely 
decoupled from the positive-parity band, i.e., the octupole
vibrational regime,  associated with the $\beta_3$-softness
of the  potential, sets in. The predicted excitation energies of 
the negative-parity states are also in good agreement with 
the experimental data though the former somewhat overestimate
the latter, in particular around $N=136$. In the case of  the 
lightest isotopes $^{218}$Ra and $^{220}$Th, the predicted 
excitation energies for both parities are too high. This 
may be a consequence of the reduced IBM space employed 
in the calculations, which is 
is not large enough to account for the 
low-lying structures of those nuclei close to the 
$N=126$
neutron shell closure. Note also that for $^{218,220}$Ra
and $^{220,222}$Th the $1^-$ energy level is 
predicted above the $3^-$ level. In the case of 
$^{220}$Ra this contradicts the experiment.
This inversion could be, once more, the result of the limited
IBM space employed in the calculations.

The present mapped $sdf$-IBM calculations, which are based on the
Gogny-D1M EDF, are able to reproduce the observed positive- and negative-parity
excitation spectra as nicely as our previous calculations 
\cite{nomura2013oct,nomura2014} employing the relativistic DD-PC1
functional. The same is true for transition strength properties. 
This agreement confirms the robustness of the SCMF-to-IBM mapping
procedure: results and conclusions do not differ at the qualitative 
(and most of the time, quantitative) level, regardless
of whether relativistic or non-relativistic energy density functional is
employed as the microscopic input.

The probability amplitudes of the 
$f$-boson components in the IBM wave functions corresponding to
even-spin positive-parity  and odd-spin 
negative-parity yrast states in $^{218-238}$Ra and $^{220-240}$Th, are 
plotted in Fig.~\ref{fig:wf} as functions of the spin $I$.
The amplitudes are computed as expectation values $\braket{\hat n_f}$ of the 
$f$-boson number 
operator $\hat n_f$ Eq.~(\ref{eq:f}) in the IBM wave  functions.
For all the studied  isotopes, at low spins $I^{\pi} \leqslant 4^+$, the fraction 
of the $f$-bosons in the positive-parity states is rather low. However, for spins 
$I^{\pi} \geqslant 6^+$ the contribution from the $f$-boson components increases
in nuclei with neutron numbers $130\leqslant N\leqslant 136$.
A similar observation applies to  negative-parity states. As can be seen from
panels (c) and (d) of the figure, the $f$-boson contributions become 
significant for  $I^{\pi} > 7^-$. For both parities and isotopic chains, the
$f$ bosons  play a major role  up to $N\approx 140$ 
even at low spins, i.e., the average value $\braket{\hat n_f}$ tends 
to be larger for lighter isotopes and becomes much smaller 
without significant changes for heavier isotopes. For the lighter
isotopes the mixing of different configurations in the $sdf$-IBM 
states is pronounced.

\subsection{Possible alternating-parity band structure}
\label{alternating-parity-band}

As a more quantitative measure of the extent to which
the predicted positive- and negative-parity bands
resemble  alternating parity bands, we 
have considered the quantity
\begin{align}
\label{alter-quant}
S(I) = E(I+1) + E(I-1) - 2E(I), 
\end{align}
where $E(I)$ represents the excitation energy of the 
$I=0^+$, $1^-$, $2^+$, $\ldots$ yrast states. In the limit 
of an ideal alternating parity band, this quantity
goes to zero. The quantity $S(I)$ is depicted in 
Fig.~\ref{fig:altern} as a function of the spin $I$. 
For most of the isotopes in both chains, the $S(I)$ values  exhibit an odd-even staggering 
pattern. This staggering pattern is less pronounced for  $N \approx 136$ reflecting 
that the negative-parity band 
becomes particularly low in energy and forms an approximate alternating-parity
structure with the positive-parity ground-state band.
For $N\geqslant 138$, the staggering is even more pronounced 
indicating that the positive- and negative-parity bands 
are decoupled from each other, a typical 
octupole vibrational feature associated 
with the $\beta_3$-softness of the potential.

\subsection{Transition strength properties}

For the computation of the reduced transition probabilities, we have employed
the  
quadrupole and octupole transition operators: 
\begin{align} \label{tran-op}
\hat T^\mathrm{E2} = e_\mathrm{B}^{(2)}\hat Q_2, 
\quad
\hat T^\mathrm{E3} = e_\mathrm{B}^{(3)}\hat Q_3
\end{align}
where $e_\mathrm{B}^{(\lambda)}$'s are effective charges and 
\begin{align}
&\hat Q_2 = s^\+ \tilde d + d^\+ s + \chi_{dd}' [d^\+\times \tilde d]^{(2)} + \chi_{ff}' [f^\+ \times \tilde f]^{(2)} \\
\label{eq:e3}
&\hat Q_3 = s^\+ \tilde f + f^\+ s + \chi_{df}' [d^\+ \times \tilde f + f^\+ \times \tilde d]^{(3)}.
\end{align}
The quadrupole and octupole transition operators Eq.~(\ref{tran-op})
have the same form as the ones in the Hamiltonian 
Eqs.~(\ref{eq:sd}) to (\ref{eq:sdf}) but with new parameters
$\chi_{dd}'$, $\chi_{ff}'$, and $\chi_{df}'$. The effective charges $e_\mathrm{B}^{(\lambda)}$'s 
are determined so that the intrinsic quadrupole (octupole) 
moment in the IBM, obtained as the expectation value of the 
operator $\hat T^{\mathrm{E}\lambda}$ in the coherent state 
at the minimum of the PES \cite{IBM} is equal to the  
Gogny-HFB one. Introducing the 
bosonic deformation 
parameters 
$\bar\beta_\lambda =C_\lambda\beta_\lambda$, corresponding to the minimum of 
the PES, we obtain the following equations 
\begin{align}
\label{eq:eb-1}
& e_\mathrm{B}^{(2)}
\frac{N_\mathrm{B}(2\bar\beta_2 - \sqrt{\frac{2}{7}}\chi_{dd}'\bar\beta_2^2 - \frac{2}{\sqrt{21}}\chi_{ff}'\bar\beta_3^2)}
{1+\bar\beta_2^2 + \bar\beta_3^2}
=Q_{20}^\mathrm{min} 
\end{align}
\begin{align}
\label{eq:eb-2}
& e_\mathrm{B}^{(3)}
\frac{N_\mathrm{B}\bar\beta_3(1 - \frac{2}{\sqrt{15}}\chi_{df}'\bar\beta_2)}
{1+\bar\beta_2^2 + \bar\beta_3^2}
=Q_{30}^\mathrm{min}. 
\end{align}
For the parameter $\chi_{dd}'$ we have adopted the value 
$\chi_{dd}'=-\sqrt{7}/2$ obtained in the SU(3) limit \cite{IBM}. On the other hand, we have 
taken the averages  $\chi_{ff}'=1.5$ and  $\chi_{df}'=-1.6$ of the 
$\chi_{ff}$ and $\chi_{df}$  values employed for the Hamiltonian, respectively. 
The effective charges $e_\mathrm{B}^{(2)}$ and $e_\mathrm{B}^{(3)}$ have 
been further multiplied by the scale factors $s_1$ and $s_2$, respectively.
The scale  factor $s_1$ is assumed to take the form 
$s_1=1.55/(9.3-0.3N_\mathrm{B})$, in order to reproduce the experimental 
systematic of the $B(\mathrm{E2}; 2^+_1\to 0^+_1)$ values.
The boson-number dependence in the denominator of 
$s_1$ has been introduced so that the computed 
$B$(E2; $2^+_1\to 0^+_1$)  is not too large 
for $N=150$
isotopes (close to the 
neutron mid-shell $N=154$). On the other hand, we have considered
$s_2=0.33$ so that an overall agreement with the systematic of the 
experimental $B(\mathrm{E3}; 3^-_1\to 0^+_1)$ values is obtained.
 
The effective charges $e_\mathrm{B}^{(2)}$ and $e_\mathrm{B}^{(3)}$ 
(in $\sqrt{\mathrm{W.u.}}$ units) Eqs.~(\ref{eq:eb-1})
and (\ref{eq:eb-2}), are plotted 
in panels (k) and (l) of Fig.~\ref{fig:parameter} as functions
of the neutron boson number $N_\nu$. The effective charge 
$e_\mathrm{B}^{(2)}$ increases smoothly with the neutron number
while  $e_\mathrm{B}^{(3)}$ exhibits a parabolic behavior
with a maximum at $N_\nu\approx 5$ that corresponds 
to  neutron numbers $N\approx 136$ at which the 
most pronounced octupole deformations are found.

The electric dipole (E1) mode is yet another characteristic property 
of pear-shaped nuclei. 
In the $sdf$-IBM framework, the E1 operator reads 
\begin{align}
\label{eq:e1}
\hat T^\mathrm{E1} = e_\mathrm{B}^{(1)} (d^\+\times\tilde f + f^\+\times\tilde d)^{(1)},
\end{align} 
with the E1 effective charge $e_\mathrm{B}^{(1)}$. We have 
taken $e_\mathrm{B}^{(1)}=0.0277$ $e\cdot$b$^{1/2}$
in order to reproduce the experimental $B(\mathrm{E1}; 1^-_1\to 0^+_1)$ 
value for 
$^{222}$Ra.

The predicted $B$(E2; $2^+_1\to 0^+_1$), $B$(E3; $3^-_1\to 0^+_1$), 
and $B$(E1; $1^-_1\to 0^+_1$) transition rates are 
compared in Fig.~\ref{fig:transitions} with the available 
experimental data. The increase  in the $B$(E2) values
(panels (a) and (b))
correlates well with the increase in quadrupole collectivity
along the studied isotopic chains. The $B$(E3) strengths
(panels (c) and (d))
display a parabolic behavior, similar to the one obtained
for the excitation energies of negative-parity states, with 
a maximum around the neutron number $N=136$.

The $B$(E1; $1^-_1\to 0^+_1$) 
strengths
(panels (e) and (f)) increase smoothly.
The predicted $B$(E1) values reproduce
the reasonably well the experimental ones for 
$^{218-222}$Ra and $^{220-226}$Th. However, the 
calculations are not able to account 
for the experimental   $B$(E1) values  in  
$^{224}$Ra and $^{228,230}$Th.
Here, one should keep
in mind that E1 transitions are less collective in nature and very sensitive to the
occupancy of high-$j$ orbitals around the Fermi surface \cite{EGIDO1990,EGIDO1992}. Due 
to this sensitivity to single particle properties, specific details of  E1 transitions may be, at least for some nuclear systems, out
of reach for the IBM description (based on collective nucleon pairs) employed 
in this study. Phenomenological IBM studies 
(see, for example, Refs.~\cite{KUSNEZOV1988,zamfir2001,zamfir2003,spieker2015}) 
have often considered the dipole 
$L=1^-$ ($p$) boson to effectively describe E1 transitions. However, such
a boson has not been included in this work since its microscopic 
origin is less clear than for the $s$, $d$, and $f$ bosons.

\subsection{Transition quadrupole and octupole moments}

The quadrupole $Q_{2}(I\to I-2)$ as well as the  octupole 
$Q_{3}(I\to I-3)$ and $Q_{3}(I\to I-1)$ 
moments, obtained from the reduced matrix elements
$\braket{I-2 \| \hat T^\mathrm{E2} \| I}$, $\braket{I-3 \| \hat T^\mathrm{E3} \| I}$, 
and $\braket{I-1 \| \hat T^\mathrm{E3} \| I}$, 
are often considered as signatures of quadrupole and octupole collectivity. 
Those transition multipole ($\lambda=2,3$) moments can be expressed as: 
\begin{align}
\sqrt{2I+1}\sqrt{\frac{2\lambda+1}{16\pi}}(I\lambda 00|I'0)Q_\lambda(I\to I')
=\braket{I' \| \hat T^{\mathrm{E}\lambda} \| I} , 
\end{align}
where  $(I\lambda 00|I'0)$ denotes a Clebsch-Gordan coefficient.
These quantities have been computed for the in-band E2 transitions within 
$K^{\pi}=0^+$ and $K^{\pi}=0^-$ bands with $| I-I' |=\Delta I = 2$, and 
for the $\Delta I=3$ and $\Delta I=1$ E3 transitions between the 
$K^{\pi}=0^+$ and $K^{\pi}=0^-$ states. 
They have been computed up to the spin $I^{\pi}=8^+$.

%
%
\begin{table}[!htb]
\begin{center}
\caption{\label{tab:ra224} Theoretical and experimental $B$(E2), 
 $B$(E3), and $B$(E1) transition rates (in Weisskopf units) for
 $^{224}$Ra. Experimental values are taken from 
 Ref.~\cite{gaffney2013}. For comparison,  results based on  the relativistic 
 DD-PC1 EDF \cite{nomura2014} have also been included in the table. 
 All transitions, exception made of  $B({\textnormal{E2}};2^{+}_{2}\rightarrow 0^{+}_{})$, are 
 between yrast states. }
 \begin{ruledtabular}
 \begin{tabular}{lccc}
  & Experiment & Theory & Ref.~\cite{nomura2014} \\
\hline
$B({\textnormal{E2}};2^{+}_{}\rightarrow 0^{+}_{})$ & 98$\pm$3 & 109 & 109 \\
$B({\textnormal{E2}};3^{-}_{}\rightarrow 1^{-}_{})$ & 93$\pm$9 & 81 & 71 \\
$B({\textnormal{E2}};4^{+}_{}\rightarrow 2^{+}_{})$ & 137$\pm$5 & 151 & 152 \\
$B({\textnormal{E2}};5^{-}_{}\rightarrow 3^{-}_{})$ & 190$\pm$60 & 103 & 97 \\
$B({\textnormal{E2}};6^{+}_{}\rightarrow 4^{+}_{})$ & 156$\pm$12 & 154 & 159 \\
$B({\textnormal{E2}};8^{+}_{}\rightarrow 6^{+}_{})$ & 180$\pm$60 & 138 & 153 \\
$B({\textnormal{E2}};2^{+}_{2}\rightarrow 0^{+}_{})$ & 1.3$\pm$0.5 & 4.9 & 0 \\
$B({\textnormal{E3}};3^{-}_{}\rightarrow 0^{+}_{})$ & 42$\pm$3 & 50 & 42 \\
$B({\textnormal{E3}};1^{-}_{}\rightarrow 2^{+}_{})$ & 210$\pm$40 & 86 & 85 \\
$B({\textnormal{E3}};3^{-}_{}\rightarrow 2^{+}_{})$ & $<$600 & 57 & 46 \\
$B({\textnormal{E3}};5^{-}_{}\rightarrow 2^{+}_{})$ & 61$\pm$17 & 85 & 61 \\
$B({\textnormal{E1}};1^{-}_{}\rightarrow 0^{+}_{})$ & $<5\times 10^{-5}$ & 4.8$\times 10^{-3}$ & 2.0$\times 10^{-3}$ \\
$B({\textnormal{E1}};1^{-}_{}\rightarrow 2^{+}_{})$ & $<1.3\times 10^{-4}$ & 5.9$\times 10^{-4}$ & $1.1\times 10^{-3}$ \\
$B({\textnormal{E1}};3^{-}_{}\rightarrow 2^{+}_{})$ & $3.9^{+1.7}_{-1.4}\times 10^{-5}$ & 1.4$\times 10^{-2}$ & 3.7$\times 10^{-3}$ \\
$B({\textnormal{E1}};5^{-}_{}\rightarrow 4^{+}_{})$ &  $4^{+3}_{-2}\times 10^{-5}$ & 2.4$\times 10^{-2}$ & 5.0$\times 10^{-3}$ \\
$B({\textnormal{E1}};7^{-}_{}\rightarrow 6^{+}_{})$ & $<3\times 10^{-4}$ & 3.5$\times 10^{-2}$ & $5.8\times 10^{-3}$ 
 \end{tabular}
 \end{ruledtabular}
\end{center} 
\end{table}

The transition quadrupole and octupole moments, obtained for
$^{218-238}$Ra and $^{220-240}$Th, are shown in 
Fig.~\ref{fig:mom} as functions of the spin $I$. 
The quadrupole moments (panels (a) and (b)) remain rather constant 
with spin although a certain staggering pattern is observed.
In the case of the octupole moments, depicted 
in panels (c) to (f) of the figure, the lightest isotopes display
rather irregular patterns with spin. However, the amplitudes 
of the oscillations become smaller for $130 \leqslant N \leqslant 136$, i.e., as
one approaches 
stable octupole deformation. On average the computed $Q_3$ moments, for 
both the $\Delta I=1$  and $\Delta I=3$ transitions, are 
around 2000 $e\cdot$fm$^3$.

\subsection{Low-energy excitation spectra, reduced 
transition probabilities and reduced matrix elements
for selected Ra isotopes}

In what follows, the  low-energy excitation spectra predicted 
for $^{224,226}$Ra are discussed in detail to further examine
the predicted power of the employed IBM framework based 
on the Gogny-D1M EDF. The energy spectrum obtained for 
$^{224}$Ra  
is compared with the experimental one \cite{gaffney2013}
in the top panel of Fig.~\ref{fig:ra224}.
The ground-state $K^{\pi}=0^+$ band is   
reproduced reasonably well by the calculations up to  $I^{\pi}=6^+$. 
However, for  $I^{\pi} \geqslant 8^+$ the predicted band looks stretched as
compared with the experiment. As can be seen from Fig.~\ref{fig:wf}, in
the case of $^{224}$Ra, the $f$-boson content of  states with $I^{\pi} \leqslant 8^+$
is $\braket{\hat n_f}\approx 0.7$ while for $I^{\pi} \geqslant 10^+$ the 
$f$-boson content turns out to be  $\braket{\hat n_f}\approx 1.5$.
For the  $K^{\pi}=0^-$ band, the $1^-_1$ (bandhead)  state is higher 
in energy than the experimental one, although features such as the moment of inertia 
and energy spacing agree well with the experiment. Up to $I^{\pi}=7^-$ the $f$-boson content
of the band is $\braket{\hat n_f}\approx 1.4$ while for $I^{\pi}=9^-$
more $f$-bosons start to play a role, i.e., $\braket{\hat n_f}\approx 3.0$.
Alternating parity doublets are visible, in both the theoretical and experimental
spectra, from $I^{\pi} = 5^-$. The predicted non-yrast $0^+_2$ and $2^+_2$ states (above 1 MeV)
have also been included in the figure. These states have a double octupole phonon nature
with $\braket{\hat n_f}\approx 2$.
In the bottom panel of Fig.~\ref{fig:ra224}, we have also plotted 
the energy spectrum obtained for 
$^{226}$Ra. This spectrum compares slightly better with the 
experiment than in the case of $^{224}$Ra. Here, the 
change in the structure of states in the $K^{\pi}=0^+$
($K^{\pi}=0^-$) band is less pronounced with 
$\braket{\hat n_f}\approx 0.8-0.9$  ($\braket{\hat n_f}\approx 1.5-1.9$ )
up to $I^{\pi}=16^+$ ($I^{\pi}= 11^-$).
Similar results are found for $^{226,228}$Th.

%
%
\begin{figure*}[htb!]
\begin{center}
\includegraphics[width=\linewidth]{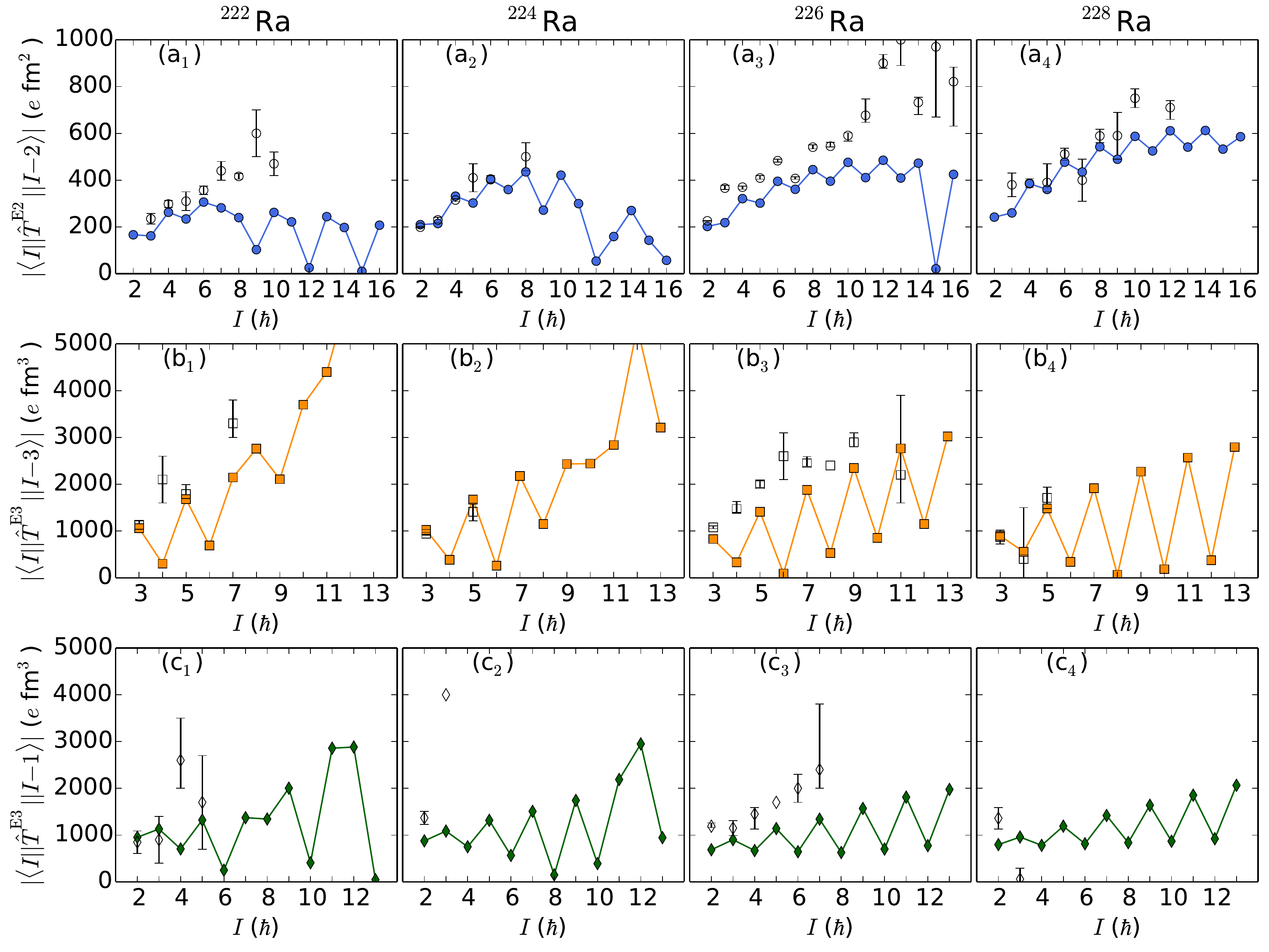}
\caption{(Color online) 
Reduced matrix elements 
$|\braket{I-2 \| \hat T^\mathrm{E2} \| I}|$, 
$|\braket{I-3 \| \hat T^\mathrm{E3} \| I}|$, 
and $|\braket{I-1 \| \hat T^\mathrm{E3} \| I}|$
for $^{222,224,226,228}$Ra. 
Experimental data are taken from Refs.~\cite{butler2020a} ($^{222,228}$Ra), 
\cite{gaffney2013} ($^{224}$Ra), 
and \cite{WOLLERSHEIM1993261} ($^{226}$Ra).
Theoretical values are represented by filled symbols  
connected by  lines. Experimental data are shown as open symbols with 
error bars. 
Experimental values without error bars 
represent upper limits \cite{gaffney2013,WOLLERSHEIM1993261} .
} 
\label{fig:ra-rme}
\end{center}
\end{figure*}
 
The $B$(E2) and $B$(E3) transition rates obtained for $^{224}$Ra are shown
in Table~\ref{tab:ra224}. We observe a very reasonable agreement with the 
corresponding experimental values. The only exceptions are the $B$(E2; $5^-\to 3^-$) and $B$(E3; $1^-\to 2^+$)
transitions which are underestimated by a factor of two to three.
In addition, we have also included in the table results from previous 
IBM calculations based on the relativistic DD-PC1 EDF \cite{nomura2014}.
As can be seen, both (mapped) IBM calculations provide rather similar predictions
for the 
$B$(E2) and 
$B$(E3) rates. The $B$(E1; $1^-\to 2^+$) values obtained in the 
present study compare slightly better with the experiment. However, other 
E1 transition strengths are larger than the ones obtained in Ref.~\cite{nomura2014}
typically by one order of magnitude and overestimate the experiment \cite{gaffney2013}   
by a factor from $10^2$ to $10^3$.

Finally, let us have a look on the reduced matrix elements 
$|\braket{I-2 \| \hat T^\mathrm{E2} \| I}|$, 
$|\braket{I-3 \| \hat T^\mathrm{E3} \| I}|$, 
and $|\braket{I-1 \| \hat T^\mathrm{E3} \| I}|$
in the case of $^{222-228}$Ra for which experimental data 
are available \cite{gaffney2013,WOLLERSHEIM1993261,butler2020a}.
They are depicted in Fig.~\ref{fig:ra-rme}  as functions of 
$I$. The predicted E2 matrix elements 
(panels (a$_{1}$) to (a$_{4}$))
increase with spin
and agree reasonably well with the experimental ones. For some 
of the studied nuclei, the E2 matrix elements are almost 
zero at high spins (for example, at $I=12^+$ for $^{224}$Ra and 
at $I=15^-$ for $^{226}$Ra). This is  probably due to
band mixing effects occurring in the  high-spin regime, as can 
be expected from the structural changes in the corresponding 
wave functions (see, Fig.~\ref{fig:wf}). The $|\braket{I-3 \| \hat T^\mathrm{E3} \| I}|$ 
(panels (b$_{1}$) to (b$_{4}$)) and 
$|\braket{I-1 \| \hat T^\mathrm{E3} \| I}|$ (panels (c$_{1}$) to (c$_{4}$))
matrix elements also increase as functions of $I$. However, they 
exhibit a pronounced staggering even at low spin that contradicts 
the pattern observed in the available experimental data. A similar 
staggering has also been obtained in previous IBM studies
\cite{nomura2014,zamfir2001}. It has been concluded, within the framework of the
phenomenological $spdf$-IBM model \cite{zamfir2001}, that at least  
$3pf$ bosons ($n_p + n_f =3$)  are required to account for the experimental systematic of the 
reduced E1 matrix elements  that linearly increase with spin.
It would be interesting to examine whether  the 
inclusion of the $p$-boson degree of freedom 
can also improve the systematic of the E3 transitions in the (mapped) IBM framework. 
Another possible remedy for the  staggering problem observed in the E3 and E1 transition 
matrix elements within the $sdf$-IBM framework could be to consider higher-order terms in 
the corresponding transition operators \cite{barfield1989} (see, Eqs.~(\ref{eq:e3}) and (\ref{eq:e1})).

\section{Summary\label{sec:summary}}

In this paper, we have considered the quadrupole-octupole coupling and  
collective excitations in the even-even actinides $^{218-238}$Ra and 
$^{220-240}$Th due to the renewed experimental interest in the region. 
The constrained Gogny-D1M HFB approach has been employed to obtain 
(axially symmetric) quadrupole-octupole SCMF-PESs. The SCMF-PESs have 
been mapped onto the corresponding IBM-PESs using the  expectation 
value of the $sdf$-IBM Hamiltonian in the boson condensate state. The 
strength parameters of the bosonic Hamiltonian have been determined via 
this mapping procedure. The wave functions resulting from the 
diagonalization of the (mapped) $sdf$-IBM Hamiltonian have been used to 
compute octupole-related quantities such as, for example, both 
positive- and negative-parity excitation spectra and transition 
strengths.


The SCMF-PESs are rather soft along the $\beta_{3}$-direction.
A global mean-field reflection-asymmetric minimum emerges 
at $N=132$ (i.e., for $^{220}$Ra and $^{222}$Th). For both isotopic
chains, the most pronounced octupole deformation effects are found 
at $N=136$ (i.e., for $^{224}$Ra and $^{226}$Th). This agrees well
with the experimental findings of
stable pear-like shapes for this particular neutron number. The 
octupole deformed minimum becomes less prominent with increasing 
neutron number and disappears from $N=142$ 
(i.e., for $^{230}$Ra and $^{232}$Th) onward. These  features 
are also found in the mapped $sdf$-IBM-PESs which nicely 
reproduce the basic topology of the fermionic PESs around the 
global minima. 

The spectroscopic properties, resulting from
the diagonalization of the $sdf$-IBM Hamiltonian, have been studied
in detail. Within this context a parabolic behavior,  centered 
around the nuclei $^{224}$Ra and $^{226}$Th, has been found 
for the low-lying negative-parity spectra and the $B$(E3; $3^+_1\to 0^+_1$)
reduced transition probabilities. For isotopes in the neighborhood
of $N=136$, an approximate alternating-parity 
band structure has been found. Octupole-related properties
have been analyzed in detail for $^{222,224,226,228}$Ra. The calculations 
reproduce reasonably well the trends observed in the data available from Coulomb excitation 
experiments. However, the fact that the calculations 
cannot account for the correct systematic of the 
$B$(E1; $1^+_1\to 0^+_1)$ rates  and/or 
the E3 transition matrix elements suggests that  improvements, such as the 
inclusion of dipole $p$ bosons, are still required in the employed mapping procedure.

From the comparison of the results obtained in this work with the 
available experimental data as well as with previous (mapped) IBM 
calculations based on the relativistic mean-field approximation 
\cite{nomura2013oct,nomura2014}, we conclude 
that the trends predicted for the studied nuclei are  
independent of the underlying microscopic input, i.e., they are robust. 
Given the predictive power and computational advantages of the 
mapping procedure together with the IBM, studies of octupolarity 
in odd-mass actinides and heavier nuclear systems appear as our next 
plausible steps.

\begin{acknowledgments}
This work has been supported by the Tenure Track Pilot Programme of 
the Croatian Science Foundation and the 
\'Ecole Polytechnique F\'ed\'erale de Lausanne, and 
the Project TTP-2018-07-3554 Exotic Nuclear Structure and Dynamics, 
with funds of the Croatian-Swiss Research Programme. The  work of LMR 
was supported by Spanish Ministry of Economy and Competitiveness (MINECO) 
Grant No. PGC2018-094583-B-I00.
This work has been partially supported by the Ministerio de Ciencia e
 Innovaci\'on (Spain) under projects number PID2019-104002GB-C21, by the
 Consejer\'{\i}a de Econom\'{\i}a, Conocimiento, Empresas y Universidad
 de la Junta de Andaluc\'{\i}a (Spain) under Group FQM-370, by the
 European Regional Development Fund (ERDF), ref.\ SOMM17/6105/UGR, and
 by the European Commission, ref.\ H2020-INFRAIA-2014-2015
 (ENSAR2). Resources supporting this work were provided by the CEAFMC
 and the Universidad de Huelva High Performance Computer (HPC@UHU)
 funded by ERDF/MINECO project UNHU-15CE-2848. 
\end{acknowledgments}

\bibliography{refs}

\end{document}